\input harvmac
\input epsf
\Title{\vbox{\hbox{HUTP-98/A029}}}
{\vbox{\hbox{\centerline{F-Theory with Quantized Fluxes}}
\vskip .3in
\hbox {\centerline{ }}}}
\vskip .3in
\centerline{Michael Bershadsky$^1$, Tony Pantev$^2$  and
Vladimir Sadov$^1$}
\vskip .2in
\centerline{\it $^1$Lyman Laboratory of Physics, Harvard
University, Cambridge, MA 02138}
\vskip .2in
\centerline{\it $^2$Department of Mathematics, University of
Pensylvania, Philadelphia, PA 19104}

\vskip .3in
\centerline{\bf Abstract}
We present evidence that the CHL string in eight dimensions is dual to 
F-theory compactified on elliptic K3 with a $\Gamma_{0}(2)$ monodromy group.
The monodromy group $\Gamma_{0}(2)$ allows one to turn on the flux of 
an antisymmetric two form along the base. The $B_{\mu \nu}$ 
flux is quantized and therefore the moduli space of the CHL string is 
disconnected from the moduli space of F-theory/Heterotic strings
(as expected). The non-zero $B_{\mu \nu}$ 
flux obstructs certain deformations restricting the 
moduli of elliptic K3 to a 10 dimensional moduli space. 
We also discuss how one can reconstruct the gauge groups from the 
elliptic fibration structure. 

\vskip .2in
\Date{March 8, 1998}

\newsec{Introduction}

F-theory appears to be a very powerful tool in analyzing the
non perturbative aspects of the heterotic string compactifications
\ref\MV{D. Morrison and C. Vafa, {\it Compactifications of F-theory on
Calabi-Yau
threefolds-
I, II}, Nucl. Phys. {\bf B 473} (1996) 74; {\it ibid}  {\bf B 476} (1996)
437.}. 
The $E_8 \times E_8$  and the $SO(32)$
heterotic string theories are equivalent to each other below ten
dimensions.
By now we have a very clear understanding of the relation between 
heterotic strings and F-theory compactifications.

Another interesting construction is the CHL string
which exists only for $d \leq 9$ \ref\cp{S.Chaudhuri, J.Polchinski,
{\it Moduli space of CHL strings,} Phys.Rev. D52 (1995) 7168, 
hep-th/9506048.} 
\ref\ss{A.Sen, J.Schwarz, {\it Type IIA
Dual of the Six-Dimensional CHL Compactification,} Phys. Lett. B357
(1995) 323, hep-th/9507027.}.
The CHL string is defined in 9 dimensions as a certain  orbifoldization  
of $E_8
\times E_8$ heterotic string. The orbifold group permutes the two
$E_8$'s and acts
on $S^1$ by a shift.
In 8 dimensions one can  also discuss CHL either as a $Spin(32)/({\bf  
Z}/2)$
heterotic string with no zero ${\bf Z}/2$ flux.
This ${\bf Z}/2$ flux according to Witten  \ref\toroidal{E.Witten, {\it
Toroidal Compactification Without Vector Structure,} hep-th/9712028.}
measures the obstruction to ``vector structure".

One can also discuss the CHL string as a $E_8 \times E_8$
heterotic string with nontrivial bundle twisted by an
external ${\bf Z}/2$ automorphism exchanging two $E_8$ \toroidal\
\ref\ler{W. Lerche, C. Schweigert, R. Minasian and S. Theisen,  
{\it A Note on the Geometryof CHL Heterotic String}, hep-th/9711104.}.
In contrast with the heterotic string  the CHL strings may have
unbroken non-simply laced gauge groups below 9 dimensions.
Some aspects of 8 dimensional compactifications were considered in
\toroidal\ where in particular the dual F-theory is investigated in
the isotrivial limit.

The dual description of CHL string in 6 dimensions and below in
terms of type
II strings was developed by A.Sen and J.Schwarz in 
\ss.  In   
the second
section we review their construction from a slightly different   
angle.  It
appears that the construction of \ss\ can be pushed up to 7
dimensions but the
8 dimensional generalization is not so straightforward.

In the third section we discuss our proposal for a F-theory dual   
to the CHL
string (in 8 dimensions). We claim that the CHL string is dual to   
a F-theory on an 
elliptic K3 with non-zero
flux of an antisymmetric 2-form $B$ through the
sphere. The value of the flux is quantized and it is fixed to be equal to 
${1 \over 2} \omega$ \foot{Only the cohomology class of the
antisymmetric 2-form $B_{\mu \nu}$ has a physical meaning. Here $\omega$ 
denotes the Kahler class of the sphere. }.  The presence of this flux
freezes 8 of the
moduli of the elliptic K3, leaving only a 10 dimensional
subspace. On this
subspace the monodromy group reduces from $SL(2, {\bf Z})$ to
$\Gamma_{0}(2)$.
This is the biggest subgroup of $SL(2 , {\bf Z})$ that keeps the
flux $B={1
\over 2} \omega$ invariant.  We will check our conjecture by
comparing the
gauge groups appearing in the F-theory and CHL string   
compactifications. The possibility of introducing a quantized $B_{\mu \nu}$
flux for certain type I backgrounds was already discussed in \ref\msp{M. Bianchi, G. Pradisi and A. Sagnotti, Nucl. Phys. B376 (1990) 365},
\ref\mas{M. Bianchi, {\it A Note on Toroidal Compactifications of the Type I Superstrings and other Superstrings Vacuum Configurations with 16 Supercharges}, hep-th/9711201}. 
Following Sen \ref\sen{A.Sen, Nucl. Phys. {\bf B475} (1996) 562 } one should expect a close relation between F-theory vacua and some type I string
backgrounds.

The mechanism of restricting the moduli to a certain sub-locus
appears to be very interesting. For example, we can turn on a flux
$B={k \over 3}
\omega$ (with $k=1,2$), restricting the moduli space of elliptic K3 to 
$\Gamma_{0}(3)$ sub-loci (in this case the moduli space turns out to have dimension six). The other interesting possibilities are 
$\Gamma_{0}(4)$ or $\Gamma_{0}(6)$ monodromy groups.

There is another interesting direction that is worth exploring
-- the relation with an M-theory compactified on a non-commutative torus 
\ref\dgn{A. Connes, M. Douglas and A. Schwarz, {\it Non commutative Geometry and Matrix Theory: Compactification on a Tori}, hep-th/9711162}
\ref\dg{M. Douglas and C. Hull, 
{\it D-branes and the Noncommutative Torus}, hep-th/9711165}.  Our 
proposal that the CHL string is 
dual to a F-theory with $B_{\mu \nu }$ flux through $P^1$ seems to be closely related to the setup considered in \dg. 
According to  \dg\ the positions of the 7-branes cease to  commute in the presence of  a $B_{\mu \nu}$ background.
It is possible that various puzzles raised in this paper can be resolved by using  non-commutative geometry.

\newsec{Low dimensional cases: 7d,  6d}

Let us start first with a six dimensional compactification ($d=3$)
(for reveiw of low dimensional compactifications see 
\ref\ozz{Shamit Kachru, Albrecht Klemm and Yaron Oz,
{\it Calabi-Yau Duals for CHL Strings}, hep-th/9712035} and references therein).
A simple generalization of  the heterotic--type II  duality
implies that the CHL string is dual to the M theory on
\eqn\mod{X_{\sigma}=\left( K3 \times S^1  \right) / (\sigma\times
(-1)), 
 }
where $\sigma$ is an involution acting  with
eight fixed points on $K3$ and $-1$ is the half period shift on
$S^1$. The action   of $\sigma\times (-1)$
has no fixed points on $K3\times S^1$ and therefore  the quotient
$X_\sigma$ is smooth and can serve as a good gravitational 
background \ss.  The number of deformations of
K3 invariant with respect to the involution $\sigma$ is equal to  12
($=$ number of hypermultiplets in 6d).
This number can either be counted by using the supergravity technique or 
just by counting directly the invariant part of the K3 cohomology.
The moduli space of this compactification coincides with the moduli   
space of the CHL string compactified on a $3$-dimensional torus and is
given by a coset space
\eqn\mod{{\cal M}_d=\Gamma  \setminus
SO(12, 4)~/  \left( SO(12) \times SO(4) \right)~,
 }
where $\Gamma$ is a certain discrete group.
The perturbative type IIA description of this compactification is   
rather obscure because the manifold $X_{\sigma}$ cannot be treated  
as an $S^1$ bundle over the quotient $K3/\sigma$: the radius of the  
would-be fiber $S^1$ is not constant over $K3/\sigma$ since it jumps  
at the orbifold points.
In the M-theory approach it is more natural to view $X_{\sigma}$ as a  
nontrivial $K3$-bundle over $S^1$ so that the limit when the circle  
gets large can be described using adiabatic arguments. In the latter  
approach the singular quotient $K3/\sigma$ never appears. Still,  
for our 8-dimensional applications we will need to understand the  
behavior of various degenerations of this quotient.

At the generic point of the moduli space \mod\ the gauge symmetry  
is broken to $U(1)^{16}$. Now let us discuss how the nonabelian  
gauge groups appear.
In this compactification we will see both simply-laced (A-D-E) and  
non-simply-laced
($B$ and $C$) gauge groups.
Recall that a simply laced group appears in M-theory   
compactified on $K3$
whenever there is a singularity on $K3$ resulting from collapsing a  
collection of 2-cycles
intersecting
each other according to an A-D-E Dynkin diagram. To find the  
simply laced gauge groups in M-theory on $X_{\sigma}$ suppose that  
there is a singular point $P$ on $K3$  {\it not fixed} by the  
involution $\sigma$, so that $\sigma(P)=P'$. The point $P'$ has to  
be singular also with the singularity of $K3$ locally isomorphic to  
that at $P$. Considering $X_{\sigma}$ as a bundle over the large  
radius circle $S^1$ we see that each fiber has (at least) two  
singularities at $P$ and $P'$ exchanged by the monodromy around the  
circle. Using adiabatic arguments the readers can easily convince  
themselves that the resulting gauge group appearing in the compactification  
of M-theory on this bundle is simply laced level $2$. This can also  
be recognized as a reformulation in the M-theory context of a  
statement in Polchinski-Chaudhuri \cp.

The mechanism responsible for the appearance of non-simply laced
gauge groups is different. Let us first note that at special points in the  
moduli space the involution $\sigma$  has invariant {\it
spheres}\foot{The existence of such spheres is a non-trivial condition
on $K3$ which will be automatically satisfied for the F-theory
compactifications dual to a CHL string. See the discussion in
Sections~3, 4 for an explanation.}. 
When such a sphere collapses to a point $K3$ develops an  
$A_1$ singularity {\it fixed} by $\sigma$.
Each such sphere passes through exactly two fixed points.
Contraction of any of these spheres leads to an enhanced gauge symmetry
$Sp(1)=SU(2)$.
This construction can be immediately generalized in order to get
$Sp(n)$ gauge
groups. Consider a collection of $2n+1$ vanishing spheres
intersecting according
$A_{2n+1}$ Dynkin diagram.  This sublattice  $A_{2n+1} \subset
H^2(K3, {\bf Z})$ maps on itself under the involution $\sigma$ in
such a way 
that two $A_n$
sublattices $\{  S_i \}$ and
$\{  S_{n+1+i} \}$  (where $i=1,\ldots, n$) are permuted
\eqn\sym{S_i \longleftrightarrow S_{2n+2-i}}
keeping  the middle sphere  $S_{n+1}$ invariant (see Fig.1).

\vskip .2in
\let\picnaturalsize=N
\def\picsize{2.4in}
\def\picfilename{A.eps}
\ifx\nopictures Y\else{\ifx\epsfloaded Y\else\fi
\global\let\epsfloaded=Y
\centerline{\ifx\picnaturalsize N\epsfxsize \picsize\fi
\epsfbox{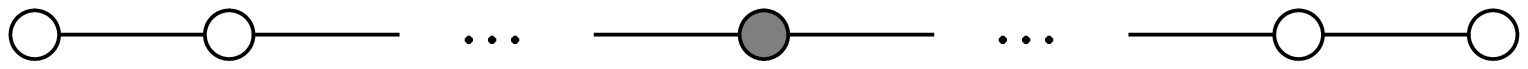}}}\fi
\vskip -.0in

\centerline{\hbox{Fig. 1}}

\noindent
One can immediately recognize that  the involution  $\sigma$
acts by an external automorphism of $A_{2n +1}$ Dynkin diagram.
This involution basically folds the Dynkin diagram on itself.
The $Sp(n)$ gauge symmetry appears after contraction of  the diagram.

To make the discussion more concrete we will use a particular  
family of elliptic $K3$ with  Weierstrass representation given by
\eqn\fkkk{
y^2 = x^3 + x f_8(z) + g_{12}(z)~,
}
where $f_8 (z)$ and $g_{12} (z)$ are polynomials on the base satisfying 
the relation $z^8 f_8 (1/z)=f_8 (z)$ and $z^{12} g_{12}
(1/z)=g_{12} (z)$.  One can immediately verify that \fkkk\ is  a  
10 dimensional family of K3's.
The involution $\sigma$ acts on \fkkk\  in a very simple way:  $y  
\rightarrow -y$ and $z \rightarrow 1/z$.

The fixed points of $\sigma$ are located in two fibers projecting  
to $z=\pm 1$ on the base. The quotient $K3/\sigma$ does not have a
Weierstrass representation, but it is still an elliptic fibration with  
two $\hat{D}_4$ fibers where the double sphere is the image of the  
elliptic fiber of $K3$.

In this model one can easily understand what happens when an invariant
sphere for $\sigma$ appears. To obtain this situation let us  
choose $f$ and $g$ such that there is an $I_2$ singular fiber at  
$z=1$. Then on the quotient the fiber is
$I^*_1$. Similarly, whenever the fiber at $z=1$ is $I_{2n}$ (the  
index can only be even because of ${\bf Z}_2$ invariance), the fiber  
on the quotient is $I^*_{n}$. The cycles in $I^*_n$ represent the  
affine Dynikin diagram $\hat{D}_{n+4}$: the four terminal points of
the graph  
represent the four orbifold points, the two cycles immediately adjacent  
to them are the ones left invariant by $\sigma$ upstairs and the $n-1$
cycles in the middle are images of $2(n-1)$ cycles upstairs
(see Fig.2).

\bigskip

\bigskip

\let\picnaturalsize=N
\def\picsize{2.0in}
\def\picfilename{Dn.eps}
\ifx\nopictures Y\else{\ifx\epsfloaded Y\else\fi
\global\let\epsfloaded=Y
\centerline{\ifx\picnaturalsize N\epsfxsize \picsize\fi
\epsfbox{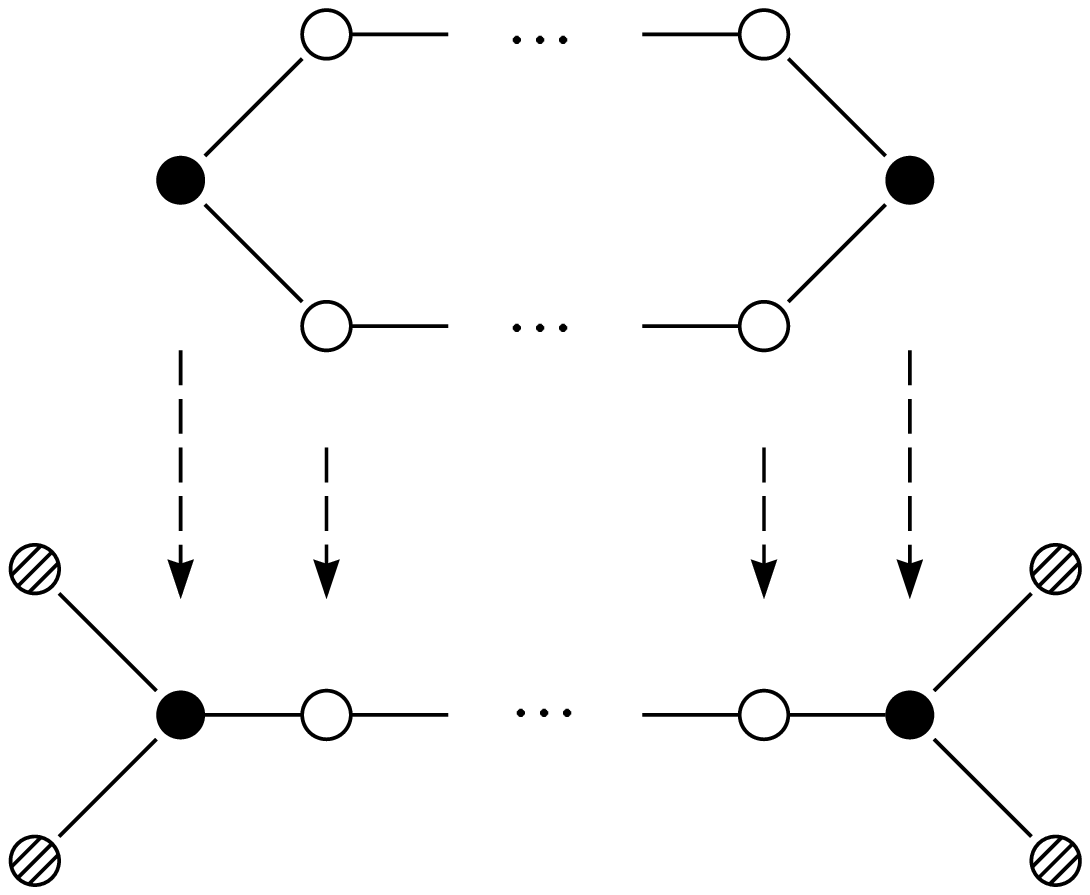}}}\fi
\vskip -.0in

\centerline{\hbox{Fig. 2}.~~
\hbox{$D_{4+n}$ singularity, four terminal cycles represent the  
orbifold points.}}

\bigskip

\noindent
The cycles upstairs represent the $\hat{A}_{2n-1}$ affine Dynkin  
diagram. The involution folds this diagram on itself keeping two  
opposite cycles fixed and attaches to each invariant cycle two more  
cycles. The singularity which corresponds to contraction of all  
these cycles except one is $(A_1)^4$ for $n=0$, $A_3\times (A_1)^2$  
for $n=1$ and $D_{n+2}\times (A^1)^2$ for $n>1$.

The following observation will be important for our discussion of  
8-dimensional compactifications in the next section. From the point  
of view of the quotient $K3/\sigma$ the above situation looks as if  
two orbifold points and an $A_n$ singularity all collide at one  
point: notice that
$A_n \times (A_1)^2 \subset D_{n+2}$, where the factor $(A_1)^2$  
describes the orbifold singularities.
To summarize, the orbifold points do not lead to massless vector  
particles
and in spite of the fact that the singularity on the quotient $K3/  
\sigma$
is $D_{n+2} \times (A_1) ^2$ the gauge group is $Sp(n)$.

In a similar fashion one can also explain the appearance of  
$SO(2n+1)$ level 1.
Consider the elliptic K3 with a degenerate fiber $I_4$ located  at  
$z=1$ and  a degenerate fiber $I_2$ located at  $z=-1$.
The zero section, which is invariant under $\sigma$  intersects one  
invariant sphere in both fibers (see Fig. 3).

\bigskip

\bigskip

\let\picnaturalsize=N
\def\picsize{2.0in}
\def\picfilename{D5.eps}
\ifx\nopictures Y\else{\ifx\epsfloaded Y\else\fi
\global\let\epsfloaded=Y
\centerline{\ifx\picnaturalsize N\epsfxsize \picsize\fi
\epsfbox{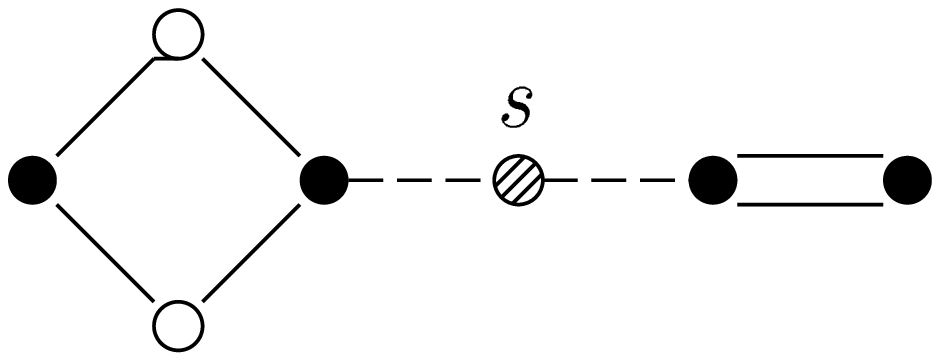}}}\fi
\vskip -.0in

\centerline{\hbox{Fig. 3}.~~\hbox{The sphere marked by $S$  
represents a zero section.}  }

\bigskip

\noindent
In turn, the invariant sphere in $I_4$ intersects two more spheres,  
which are permuted by $\sigma$.
All in all, there are five cycles representing a $D_5$ Dynkin diagram. 
The involution maps the diagram on itself  resulting in $B_4$ diagram.
Contraction of all these spheres yields $B_4=SO(9)$ gauge group.
 In order to get $SO(7)$ one has to start with $I_4$
singularity located either at $z=1$ or $z=-1$.
$SO(9)$ is the maximal $SO(2n+1)$ gauge group that can appear in  
{\it 7 dimensions}. In 6 dimensions the maximal
$SO(2n+1)$ gauge group is $SO(13)$. To see it one needs to  adjust the 
radius of the circle $S^1$.

To obtain compactifications dual to CHL string in 7d rather than in  
6d we may use a trick due to Aspinwall and Schwarz \ref\aspin{P.  
Aspinwall, {\it M-Theory Versus F-Theory Pictures of the Heterotic
String}, Adv. Theor. Math. Phys. 1 (1998) 127-147, hep-th/9707014.}.
We assume that K3 is {\it elliptic} and consider
the limit in
which the K\"{a}hler class of the elliptic fiber tends to zero. This is   
the limit in which one space dimension gets decompactified, as
described in \aspin\ in detail.
The resulting 7d theory can be described as a F-theory  
compactification on the quotient space $(K3\times S^1)/({\bf Z}/2)$. The  
other way to obtain this result is to use the duality between  
F-theory on $K3$ and heterotic string on $T^2$ in 8 dimensions and  
compactify both theories on $S^1$ with Polchinsky-Chaudhuri twist.
The discussion of gauge groups appearing in this compactification  
remains the same as in the six-dimensional case.

\newsec{F-theory with fluxes}

In this section we describe the F-theory
compactification (in 8 dimensions) dual to the CHL
string. Such F-theory compactifications were analyzed by Witten in 
\toroidal\
in the limiting locus when the elliptic fibration becomes isotrivial
(this limit was also discussed in 
\mas, \ref\jp{J. Park, Phys. Lett.  B 418 
(1998) 91 }).
We extend Witten's description to the interior of the moduli space
and confirm that even though the F-theory elliptic fibration is no
longer isotrivial the essential features of Witten's description are
retained. The extra structure of the compactification exploited here
is dictated to us by the specialization of the usual
Heterotic/F-theory duality to the CHL case. This transition is
explained in detail in the next section.

We will present strong evidence showing that CHL is dual to  
F-theory compactified on elliptic K3 with non-zero flux of an
antisymmetric field $B_{\mu\nu}$ along the base sphere.
According to the conventional wisdom, the single valued
background of antisymmetric 2-forms (NS-NS and RR) contradicts the
$SL(2,{\bf Z})$ monodromy. Luckily, there is a way out of this
contradiction when the monodromy group is smaller than $SL(2,{\bf Z})$.
The  K3 surface on which our F-theory is compactified moves in a 
10 dimensional family of  elliptic fibrations
with {\it monodromy} contained in the congruence subgroup
${\Gamma}_{0}(2)$. This monodromy group does allow (see Section~4.4 for
more details) a non-zero value of the antisymmetric 2-form. 
Before we proceed let us make some comments about the 
$B_{\mu \nu}$ backgrounds. The antisymmetric two form is defined modulo gauge transformations $B \rightarrow
B+d  \Lambda$. 
Therefore, turning the $B_{\mu \nu}$ background along the 
${\bf P}^1$ (the base of F-theory compactification) introduces
only {\it two} new real parameters -- the classes of 
$B^{NS}=b \omega$
and
$B^{RR}=b' \omega$.
Both $b$ and $b'$ take values in $S^1$. Moreover, the field
strength $H=dB$ is identically equal to zero for this 
background and therefore, locally one can gauge away the 
$B_{\mu \nu}$ field. This will become important in our further discussion.

The monodromy group 
$\Gamma_{0}(2)$ keeps one of the three half-periods
invariant. For example, realizing $\Gamma_{0}(2)$ by $SL(2, {\bf Z})$
matrices with even entries in the lower left corner we get
\eqn\gt{\pmatrix{
a & b \cr
2k & d \cr
} \left (\matrix{
1 /2 \cr
0\cr
}\right )=\left (\matrix{
a /2 \cr
k \cr
}\right )\equiv \left (\matrix{
1 /2 \cr
0\cr
}\right ) ({\rm mod} ~{\bf Z})~.}
This is the reason why one can turn on an antisymmetric 2-form without
ruining the F-theory structure,  by setting $b=1/2$ and $b'=0$.
  Moreover, the non-zero $B_{\mu \nu}$
background obstructs certain deformations of the elliptic K3, allowing  
only those compatible with $\Gamma_{0}(2)$ monodromy.

The formulation of F-theory as type IIB string is sensitive only to  
the complex moduli of elliptic manifold and it does not care
about the K\"ahler moduli. Therefore before going further we
have to choose an appropriate model for elliptic K3 with $\Gamma_{0}(2)$ 
monodromy. Let us first describe a K3 which is the Weierstrass 
model of an elliptic  fibration with  
$\Gamma_{0}(2)$ monodromy.
$\Gamma_{0}(2)$ preserves one of the non-trivial spin structures  
(half-periods) of the fiber.
That means that besides the always present section at infinity,   
the elliptic fibration has another global
section given by that half-period. The appropriate Weierstrass form
 can be written as
\eqn\model{y^2=(x- a_4(z))(x^2+x  a_4(z)  +  b_8(z))~,}
where $a_4(z),~b_8(z)$ are two polynomials of degree $4$ and $8$ on  
the base of the elliptic fibration.
It is the reducibility  of the right hand side of \model\  
that ensures the existence of the global section $y=0~, x=a_4$.
The Weierstrass equation \model\ describes a 10 parameter family of
elliptic K3 surfaces. The minimal resolution of all K3's in this
family have eight singular fibers of Kodaira type $I_{2}$ as can be
easily seen from the discriminant
\eqn\disc{\Delta=(b_8+2 a_4 ^2)^2 (4 b_8-a_4 ^2)}
The zeroes of $(b_8+2 a_4 ^2)$   determine the locations of  eight $I_2$
fibers. At the generic point in the moduli space such a K3 has eight  
$A_1$ singularities localized on the extra section $y=0, x=a_4$.

It turns out however that the moduli space of F-theory
compactifications corresponding to the CHL string is a finite
cover of the moduli space of $\Gamma_{0}(2)$ Weierstrass  models just
described. Concretely in order to recover the dual CHL compactification
from the Weierstrass model \model\ one also needs to choose 4 of the 
eight points on the base at which the $I_{2}$ fibers are located (see
Section 4.4 for an explanation). With this extra data the elliptic
curve on which the dual CHL string is compactified is recovered as the
double cover of ${\bf P}^{1}$ branched at the 4 marked points and the
twisted $E_{8}\times E_{8}$ instanton of the CHL compactification is
encoded in the complex structure of the surface \model. This situation is 
very different from the standard F-theory/Heterotic correspondence.
In the later case the elliptic curve of Heterotic compactification
(at least in the vicinity of $E_8 \times E_8$ locus) was encoded
in the elliptic fiber.

Actually from the duality with the CHL string one canonically
reconstructs (see Sections 4.3 and 4.4) a slightly different 
birational model of \model\ obtained
as follows. The minimal resolution of \model\ is a smooth elliptic K3
having eight $I_{2}$ fibers and two sections arranged in such a way
that at each $I_{2}$ fiber the two sections pass trough two different
components of the fiber. If we now contract the four components of
\model\ meeting the section $y=0, x=a_4$ at the four marked points on
the base and contract the components meeting the section at infinity
at the remaining 4 points we will get an elliptic K3 which also has
eight $A_{1}$ singularities and is the same as \model\ outside of the
$I_{2}$ fibers. In particular this new K3 has the same $j$-function 
and the
same monodromy as \model\ but is not in Weierstrass form since each of
the two sections now passes trough four of the singular points.

To make a connection between this picture (in 8 dimensions) and the
one given above in lower dimensions
let us further compactify F-theory on an extra $T^2$. As the result we
get a 6-dimensional theory which can be interpreted as M-theory on
$K3 \times S^1 /\sigma$, described in the previous section.
Let us denote the quotient $K3/\sigma$ by $X$.
As we already explained in the previous section $X$ is a singular  
K3 surface.
The surface $X$ turns out to be quite remarkable. It admits several  
elliptic fibrations two of them being particularly  
important. One elliptic fibration is inherited  from the $K3$  
upstairs (the Weierstras model for the $K3$ upstairs is given 
by \model) and will be called the {\it inherited elliptic fibration}.   
In another elliptic fibration, to be called the $\Gamma_{0}(2)$ {\it
elliptic fibration}, the monodromy group is $\Gamma_{0}(2)$  
and there are two sections (the zero section and the section at  
infinity)
with four $A_1$ singularities on top of each. Thus the surface $X$  
is exactly the elliptic K3 we chose to compactify F-theory on.

\vskip .2in
\let\picnaturalsize=N
\def\picsize{3.0in}
\def\picfilename{double.eps}
\ifx\nopictures Y\else{\ifx\epsfloaded Y\else\fi
\global\let\epsfloaded=Y
\centerline{\ifx\picnaturalsize N\epsfxsize \picsize\fi
\epsfbox{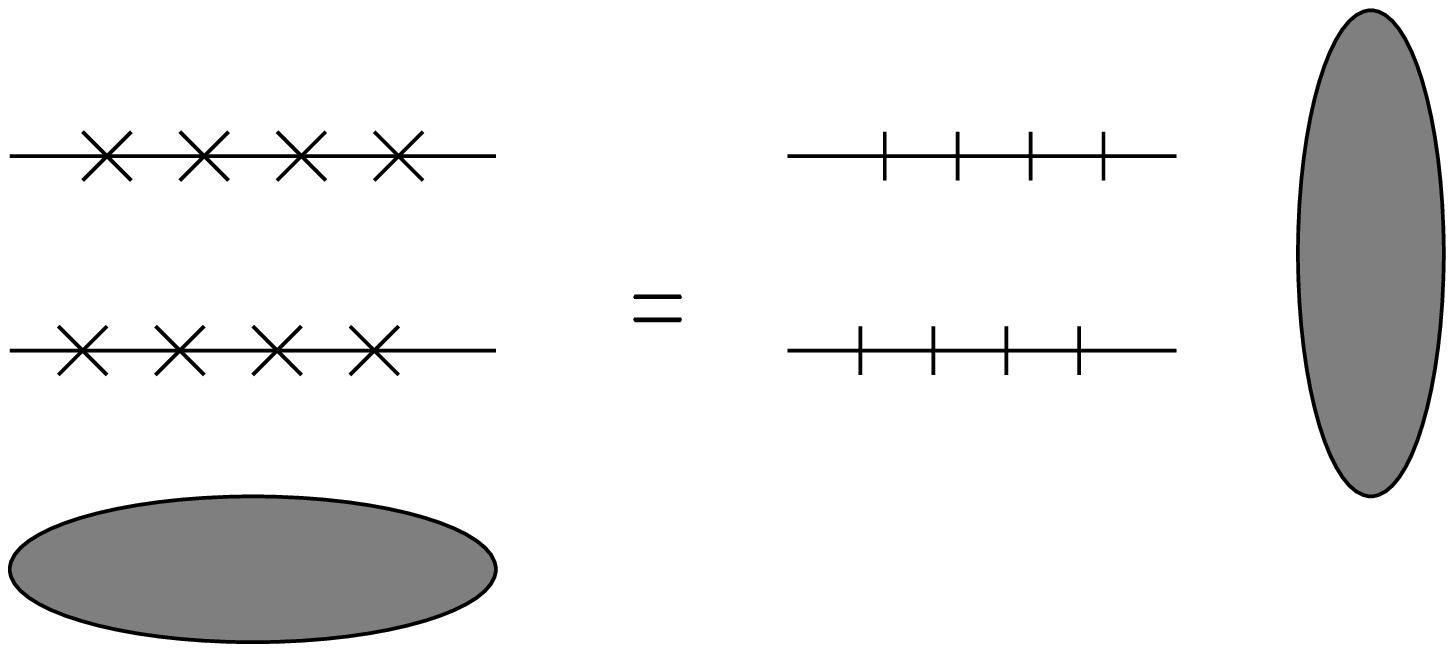}}}\fi
\vskip -.0in

\centerline{\hbox{Fig. 4}.~~\hbox{Two elliptic fibrations.}}
\vskip .2in

The existence of two elliptic structures simplifies significantly  
the identification of the points on the moduli space corresponding  
to non-abelian gauge groups.  Particularly useful is the following  
simple remark. Two sections with four $A_1$ singularities in the  
fibration with $\Gamma_{0}(2)$ are the fibers over the fixed points  
($z=\pm$ in \fkkk) in the another fibration.

Now we can turn to the discussion of gauge groups in the F-theory  
language.
The CHL string has A-D-E level 2  as well as $C_n=Sp(n)$ level 1 gauge 
groups
\foot{The maximal $SO(2n+1)$ gauge group that appear in 8  
dimensions is $SO(5)=Sp(2)$.}.
In the F-theory without the flux the gauge groups would appear at  
points in the
moduli space where elliptic fibration develops a singularity in the
elliptic fiber.
In the presence of fluxes the situation should becomes more involved.
For example, $E_6$ or $E_8$ degeneration of the elliptic fiber
cannot appear in elliptic K3 with $\Gamma_{0}(2)$ monodromy.
Still, we expect that these gauge groups appear in 8-dimensional
compactifications of the CHL string.
Moreover,  we will see that in certain cases the  presence of 
an $I_2$
singularity does not necessarily lead to a $SU(2)$ gauge group.
It seems unlikely  that these phenomena can be explained
from the point of view of local D-brane physics in the presence of
a $B_{\mu \nu}$ field. Indeed,  by choosing an appropriate gauge locally one can put the $B_{\mu \nu}$ field equal to zero
getting back the familiar D-brane picture.
There is clearly a global obstruction of getting rid of 
$B_{\mu \nu}$ field everywhere and therefore 
the absence of  the enhanced gauge 
symmetry one naively expects is forced by the global topological properties of the 
compactification\foot{Note that in general the global topology does not kill the observed gauge group completely but always reduces the naive gauge symmetry to a smaller one.}.  As explained later, geometrically this is expressed in the relative position of  the additional section and the singular fiber responsible for the appearance of the gauge symmetry. 

We suggest the following  ``strategy'' to derive the gauge groups.
As we have seen the K3 surface $X$ has multiple personalities and
disguises either as a space on which some F-theory with fluxes is 
compactified or as an object dual to the CHL string. This personality
disorder is encoded in the existence of the inherited and the
$\Gamma_{0}(2)$-elliptic fibrations on $X$ and so the idea is to compare 
the singularities that appear in these two  
elliptic fibrations. Clearly, a fiber singular in one of the fibrations
does not necessarily correspond to a singular fiber in the other 
elliptic fibration. It may correspond to a collection of singular
fibers connected by a section (or by sections). 
For simple gauge groups the corresponding singularities of the 
total space after the resolution can be localized on one 
singular fibers in {\it each} elliptic fibration.
By the nature of the 
duality construction the CHL gauge groups can be  detected by 
analyzing the 
singular fibers in the inherited elliptic fibration. 
For all gauge groups but the exceptional ones,
the Dynkin diagram of the gauge group coincides with the common part
of the singular fibers taken in  both elliptic fibrations,
with the cycles corresponding to the fixed points being dropped out.
Therefore to see
these groups in the F-theory with fluxes we need to translate
(see Section~4.5) the singularities into collision patterns for the 
$\Gamma_{0}(2)$-fibration. This can be done either directly on $X$ or on
its $\Gamma_{0}(2)$-Weierstrass model $w : Y \rightarrow {\bf P}^{1}$. In
the latter case however we need to remember the extra data of
splitting the eight $I_{2}$ fibers into two groups of four $I_{2}$
fibers each.

For example, if a Kodaira fiber of affine A-D-E type appears in the
inherited elliptic fibration in a way that does not involve collision
with the two $\hat{D}_{4}$ fibers, then when viewed in the
$\Gamma_{0}(2)$-elliptic fibration this fiber gives rise to the condition
for the appearance of an A-D-E gauge group of level 2.
In order to get the $Sp(n)$ level 1 gauge groups one has to start with
$I_n$ colliding with one of the $\hat{D}_4$ fibers. As the result we
get the following ``dictionary''.

\subsec{$Sp(n)$ gauge groups}

The $Sp(n)$ gauge groups appear as the result of collision of
two $I_2$ fibers {\it from the same section} and $nI_1$ fibers for  
$n>1$. In terms of the Weierstrass model $Y$ this just means that the
two colliding $I_{2}$ fibers belong to the same group of four.

The case $Sp(1)=SU(2)$ is special and this gauge group appears when
just two $I_2$ fibers collide with each other.
For every $n>1$ the collision $2I_2 + nI_1$ produces a $\hat{D}_{n+2}$
singular fiber in the $\Gamma_{0}(2)$-fibration $\hat{\pi} : \hat{X}
\rightarrow {\bf P}^{1}$ on the minimal resolution of $\hat{X}$. 
The corresponding singularity of $X$ (or of the Weierstrass model $Y$)
is obtained by all cycles in $\hat{D}_{n+2}$ which do not meet the
zero section of $Y$. This cycles form an ordinary $D_{n+2}$ Dynkin
diagram and the two cycles on the tail of this $D_{n+2}$ correspond to
two of the branch points of the covering $S \rightarrow X$.
The other $n$ represent the roots of $Sp(n)$.
The maximal rank that can be obtained this way is $8$.
Still, in CHL string one can get get $Sp(9)$ and $Sp(10)$. 
In this case the monodromy group further reduces
to $\Gamma(2) \in \Gamma_0(2)$.

\subsec{$SU(n)$}

Collision of $n$ fibers of type $I_1$ among themselves leads to a 
$SU(n)$ gauge group. The maximal rank that can be obtained in this way is $8$.

\subsec{$SO(2n)$ gauge groups}

The $SO(2n)$ gauge groups appear as the result of collision of
two $I_2$ fibers {\it from different sections}  and $nI_1$ fibers  
for $n>1$. In the Weiersrass model this just corresponds to the collision
of two $I_{2}$ fibers belonging to two different groups of four.
The maximal gauge group that appears this way is $SO(16)$.

\subsec{Exceptional gauge groups ($E_6$, $E_7$, $E_8$)}

The exceptional groups of the $E_{n}$ series appear when a number of
$I_{1}$ fibers in the $\Gamma_{0}(2)$-fibration collide in such a way that
the monodromy reduces to the subgroup $\Gamma(2) \subset
\Gamma_{0}(2)$ which preserves all points of order two\foot{Explicitly
$\Gamma(2) = \left\{ \left.
\pmatrix{
a & b \cr
c & d \cr} \in SL(2,{\bf Z}) \right|a, d - \, {\rm odd}, \; b, c - \, 
{\rm even}
\right\}$ }. Concretely
this means that the bisection $C \subset X$ splits into two
sections. In the Weierstrass model $Y$ given by \model\ the curve $C$
is given by the equation $y = 0 = x^{a} + a_{4}(z)x + b_{8}(z)$
and so the $\Gamma(2)$ degeneration occurs exactly when $x^{2} + 
a_{4}(z)x + b_{8}(z)$ decomposes as $x^{2} + 
a_{4}(z)x + b_{8}(z) = (x - p(z))(x - q(z))$ with $p(z)$, $q(z)$ -
general polynomials of degree four on ${\bf P}^{1}$.

After tracing (see Section 4.5 for details) the correspondence between
the singular fiber in the two elliptic fibrations one sees that the
exceptional gauge groups appear as follows:

\medskip

\noindent
$\underline{E_{6}}$: Corresponds to a $\Gamma(2)$-collision 
of six fibers of type $I_{1}$
in the $\Gamma_{0}(2)$ elliptic fibration.

\medskip

\noindent
$\underline{E_{7}}$:
Can be obtained in two ways. Either as a $\Gamma(2)$-collision of
eight fibers of type $I_{1}$ in the $\Gamma_{0}(2)$ elliptic fibration
or as a $\Gamma(2)$-collision of six $I_{1}$ fibers with two $I_{2}$
fibers coming from the two different sections in the $\Gamma_{0}(2)$ 
fibration.

\medskip

\noindent
$\underline{E_{8}}$:
Corresponds to a $\Gamma(2)$-collision of eight $I_{1}$ fibers with
two $I_{2}$ fibers coming from the two different sections in 
the $\Gamma_{0}(2)$ fibration.

\bigskip

Furthermore, it turns out (see Section~4.5) that the two mechanisms of
$E_{7}$ creation described above are not deformable to each within the
locus of deformations of $X$ preserving the $\Gamma(2)$
monodromy. On the other hand the mechanism of crating $E_{6}$ deforms
further to $E_{7}$ coming from a $\Gamma(2)$-collision of $8I_{1}$ and
the occurrence of $E_{7}$ as a $\Gamma(2)$-collision of $6I_{1} +
2I_{2}$ admits a further $\Gamma(2)$ degeneration to the creation of
$E_{8}$. 

The two ways of creating $E_{7}$ can be deformed to each other by
keeping both an $E_{7}$ fiber in the inherited elliptic fibration and
a $\Gamma_{0}(2)$ monodromy. The intermediate $\Gamma_{0}(2)$ surfaces
however are singular in codimension one and for them the $j$-invariant
of the $\Gamma_{0}(2)$-fibration becomes identically $\infty$.

\subsec{Other collisions}

Other collisions usually correspond to products of gauge groups.
For example the singularity $3I_2 + n I_1$ gives rise to the  
$Sp(n)\times Sp(1)$ gauge groups for $n \geq 2$ and $(Sp(1))^2$  
gauge group for $n=0$.
The series $4I_2 + n I_1$ corresponds to $Sp(n)\times (Sp(1))^2$ for  
$n \geq 2$ and $(Sp(1))^3$ for $n=0$.

\bigskip

\noindent
{\bf Remark.}  The A-D-C groups of ranks bigger than eight should appear in a way similar to the creation of the exceptional gauge symmetries. In other words one expects higher rank A-D-C's when  the monodromy group drops to some finite index subgroup in 
$\Gamma_{0}(2)$.

\newsec{Justification of the CHL/F-theory duality}

In this section we transplant the usual duality mechanism relating
F-theory and the Heterotic string to the situation of the
CHL/F-theory correspondence. For the convenience of the reader and to
set up notation the
relevant parts of the Heterotic/F-theory duality (see
\ref\vmtwo{D.Morrison,
C.Vafa, {\it Compactifications of F-Theory on
Calabi--Yau Threefolds -- II}, Nucl.Phys. {\bf B476} (1996) 437-469,
hep-th/9603161.},  \ref\us{M.Bershadsky, A.Johansen,
T.Pantev, V.Sadov, {\it On Four-Dimensional Compactifications of
F-Theory}, Nucl.Phys. {\bf B505} (1997) 165-201, hep-th/9701165.},
\ref\fmw{R.Friedman, J.Morgan, E.Witten, {\it Vector Bundles And F
Theory}, Commun.Math.Phys. 187 (1997) 679-743, hep-th/9701162.} and 
\ref\ronictp{R.Donagi, {\it ICMP lecture on Heterotic/F-theory
duality}, hep-th/9802093.} for more details) are reviewed in the
Appendix.

\subsec{The on-the-nose correspondence}

The CHL string can be obtained from the
$E_8 \times E_8$ Heterotic string  by a certain ${\bf Z}/2$ quotient.
The CHL involution acts by a half period shift on the torus $T^{2}$
and permutes the two $E_8$. Clearly this involution acts trivially on
$\tau$ and $\rho$ (complex and K\"{a}hler strictures of $T^2$)
and so the moduli space of vacua for CHL
compactified on $T^{2}$ can be naturally identified
\cp\  with the locally symmetric
space
\eqn\chlspace{
O(2,10;{\bf Z})\backslash O(2,10;{\bf R})/(O(2;{\bf R})\times
O(10;{\bf R}))~.
}
Intrinsically the relation of \chlspace \ with
(5.1) can be
described as follows. First use the Narain construction to write
(5.1) in a more precise form, namely as the double coset space
$$
O({\bf Nar})\backslash O({\bf Nar}\otimes {\bf R})/K
$$
where\foot{Here ${\bf U}$ denotes the two dimensional
hyperbolic lattice and ${\bf E}_{8}$ denotes the root lattice of the
group $E_{8}$.}
${\bf Nar} := {\bf U}\oplus {\bf U}\oplus {\bf E}_{8}\oplus {\bf
E}_{8}$ is the Narain lattice and $K \subset O({\bf Nar}\otimes {\bf
R})$ is a maximal compact subgroup. As usual the ${\bf U}\oplus {\bf
U}$ part of the Narain lattice is identified\foot{Since the pair
$(\tau,\rho)$ intrinsically lives in $(H^{2}(T^{2},{\bf
Z})\oplus H^{2}(T^{2},{\bf Z}))\otimes {\bf R}$} with $H^{2}(T^{2},{\bf
Z})\oplus H^{2}(T^{2},{\bf Z})$. The CHL involution which acts as a
translation by a half period on the torus and is homotopic to the
identity and so induces the identity on cohomology. Thus the CHL
involution $i : {\bf Nar} \rightarrow {\bf Nar}$ is
given explicitly by $i(u_{1},u_{2};\xi',\xi'') =  
(u_{1},u_{2};\xi'',\xi')$.
{}From that viewpoint the space \chlspace \ is naturally written as
$O({\bf Nar}^{\langle i \rangle})\backslash O(({\bf Nar} \otimes {\bf
R})^{\langle i \rangle})/({\rm maximal \; compact})$ with
${\bf Nar}^{\langle i \rangle}$ being
the sublattice of $i$-invariants. From the exact
form of $i$ one easily sees that\foot{As usual given an
Eucledian lattice $L$ and an integer $m$ we denote by $L(m)$  the
lattice isomorphic to $L$ as a free abelian group but whose bilinear
pairing is $m$ times the pairing of $L$.} ${\bf Nar}^{\langle i  
\rangle} \cong
{\bf U} \oplus {\bf U} \oplus {\bf E}_{8}(2)$ which suggests that the
dual F-theory should be compactified on a K3 surface\foot{See the
Appendix of how moduli spaces of lattice polarized K3 surfaces
naturally appear in the context of F-theory.} whose
transcendental lattice is at least isogenous to ${\bf U} \oplus {\bf
U} \oplus {\bf E}_{8}(2)$. In fact there is a well understood (see
\ref\dol{I.Dolgachev, {\it Mirror symmetry for
lattice polarized K3 surfaces}, alg-geom/9502005},
Example~9.1 and  \ref\nam{Y.Namikawa, {\it Periods of Enriques surfaces},
Math. Annalen, 270 (1985), 201-222.}) moduli
space of lattice polarized K3's with this property - the moduli space
of K3 surfaces which are universal covers of Enriques
surfaces.

Recall \ref\dolenr{F.Cossec, I.Dolgachev, {\it Enriques surfaces I},
Progress in Math. vol. 76, Birkh\"{a}user, 1989.} that an Enriques
surface is a complex surface $F$ with
$H^{0}(F,\Omega^{1}_{F}) = H^{0}(F,\Omega^{2}_{F}) = 0$ but for which 
$K_{F}^{\otimes 2} = (\Omega^{2}_{F})^{\otimes 2} = {\cal O}_{F}$. The
unramified double covering $p: S \rightarrow F$ corresponding to the
2-torsion line bundle $K_{F}$ has trivial canonical class and no
holomorphic one forms and is thus a K3 surface. Let $e : S \rightarrow
S$ be the fixed point free involution acting along the fibers of $p$.
The second cohomology
group $H^{2}(F,{\bf Z})$ is isomorphic to ${\bf U}\oplus {\bf
E}_{8}$. The pullback map gives an inclusion $p^{*} : H^{2}(F,{\bf  
Z})/{\rm
torsion} \hookrightarrow H^{2}(S,{\bf Z})$ of free abelian groups but
since $p$ is two-sheeted we get an isomorphism of Eucledian lattices 
$H^{2}(S,{\bf Z})^{\langle e \rangle} = p^{*}H^{2}(F,{\bf Z}) \cong
{\bf U}(2)\oplus  {\bf E}_{8}(2) =: M$. One can check \dolenr\ that ${\rm
Pic}(F) = H^{2}(F,{\bf Z})$ and that for the generic $F$ all algebraic
cycles on $S$ are pullback from cycles on $F$. In this way $S$
acquires a canonical structure of ${\bf U}(2)\oplus  {\bf
E}_{8}(2)$-polarized K3 surface. It can be shown \nam\ that we can
choose a marking $H^{2}(S,{\bf Z}) \cong {\bf U}\oplus {\bf
U}\oplus {\bf U}\oplus  {\bf E}_{8}\oplus {\bf E}_{8} =: \Lambda$ so
that the Enriques involution $e$ acts as  
$e(u_{1},u_{2},u_{3};\xi',\xi'') =
(-u_{1},u_{3},u_{2};\xi'',\xi')$. In particular the transcendental
lattice of a general $S$ will be isomorphic to  $\Lambda^{{\rm  
sign}(\langle
e\rangle)} =  {\bf U}\oplus {\bf U}(2)\oplus {\bf E}_{8}(2)$ and hence
the moduli space of all $S$'s will be isomorphic\foot{See the
discussion in the Appendix.} to $O(\Gamma_{M})\backslash
O(\Lambda^{{\rm sign}(\langle e\rangle)}\otimes {\bf R})/({\rm maximal
\; compact})$. Up to isogeny this space is isomorphic to \chlspace\
and this identification provides the first approximation to the
F-theory--CHL duality.

\subsec{The correspondence in the stable limit}

To reconfirm that the F-theory compactifications described in the
previous subsection are indeed dual to the CHL string it is
instructive to examine the above correspondence in the stable limit
where the complexified K\"{a}hler class of the CHL $T^{2}$ is allowed
to become infinitely large.

Similarly to the ordinary F-theory/Heterotic duality (see the
Appendix) in this regime the vacua for both the CHL string and the
F-theory can be encoded entirely in algebraic-geometric objects. For
the CHL string we have:

\medskip

\noindent
(a) An elliptic curve $E$ with a marked
origin (= the torus $T^{2}$ taken with the complex structure $\tau$).

\medskip

\noindent
(b) A point of order two $\alpha \in E$ (so that translation
$t_{\alpha} : E \rightarrow E$ by $\alpha$ corresponds to the action
of the CHL involution on $T^{2}$). We denote the quotient elliptic
curve $E/\langle t_{\alpha} \rangle$ by $Z$ and the natural projection
$E \rightarrow Z$ by $q_{\alpha}$.

\medskip

\noindent
(c) An $\alpha$-twisted $E_{8}\times E_{8}$ principal bundle $V
\rightarrow Z$.

\bigskip

To explain the notion of an $\alpha$-twisting start by recalling that
a principal $G$-bundle on a variety $B$ is a variety $P$ equipped
with a locally trivial fibration $f : P \rightarrow B$ and a right
action $P\times G \rightarrow P$ such that for every $b \in B$ the
action on the fiber $P_{b}\times G \rightarrow P_{b}$ is simply
transitive. In exactly the same way we can start not with a group $G$
but with a twisted form of $G$, i.e. a bundle of groups ${\cal G}
\rightarrow B$ which on small open sets $U \subset B$ is isomorphic to
the product of $G\times U$. Now a principal ${\cal G}$ bundle (or
${\cal G}$-torsors) will be
just a variety $P$ equipped with a locally trivial fibration $f : P
\rightarrow B$ and an action $P\times {\cal G} \rightarrow P$ which
is fiber-wise simply transitive. In the concrete situation above one
considers the bundle of groups ${\cal E}_{8}^{\alpha} \rightarrow Z$
defined as the push-forward $q_{\alpha*}\underline{E}_{8}$ of the
trivial bundle of groups $\underline{E}_{8} = E\times E_{8}$ on $E$.
The fiber of ${\cal E}_{8}^{\alpha}$ over a point $z \in Z$ is
isomorphic to the product of two copies of the group $E_{8}$ labeled
by the two points in the fiber $q_{\alpha}^{-1}(z) \subset E$.

According to our on-the-nose correspondence and the usual
F-theory/Heterotic duality the data (a), (b) and (c) will have to
correspond to a stable degeneration of a K3 cover of an Enriques
surface. In order to describe this degeneration we will have to view
the data (a), (b) and (c) as a data describing an ordinary Heterotic
string in 8 dimensions. To do that we have to recast the
$\alpha$-twisted principal bundle $V$ as an object on $E$. For this we
just pull $V$ back via the covering map $q_{\alpha} : E \rightarrow
Z$. By definition $q_{\alpha}^{*}V$ will be a principal bundle over 
$q_{\alpha}^{*}{\cal E}^{\alpha}_{8} = \underline{E}_{8}\times
\underline{E}_{8}$, i.e. an ordinary $E_{8}\times E_{8}$ bundle on
$E$. Furthermore $t_{\alpha}^{*}q^{*}_{\alpha}V \cong q^{*}_{\alpha}V$
by construction and so $q^{*}_{\alpha}V = W\times t_{\alpha}^{*}W$ for
some $E_{8}$-bundle $W$ on $E$. Conversely given an $E_{8}$-bundle $W
\rightarrow E$ we can define $V = q_{\alpha *}W = (q_{\alpha *}(W\times
t_{\alpha}^{*}W))^{\langle t_{\alpha} \rangle}$. To summarize: the CHL
data (a), (b), (c) on $Z$ is equivalent to the Heterotic string data
consisting of $E$, the involution $t_{\alpha} : E \rightarrow E$ and
an $E_{8}$ bundle $W$ on $E$. By what we have just said the
$E_{8}\times E_{8}$ bundle on $E$ that one gets in this way is just
the bundle $W\times t_{\alpha}^{*}W$. Therefore, as explained in the
Appendix, the stable degeneration of the K3 corresponding to $E$ and
$W\times t_{\alpha}^{*}W$ will have to be a union of two rational
elliptic surfaces (with sections) $S'$ and $S''$ intersecting
transversally along a
copy of $E$ sitting in each of them as an anticanonical section. The
only new feature in this picture is that since the two bundles $W$ and
$t_{\alpha}^{*}W$ are isomorphic the two rational elliptic surfaces
$S'$ and $S''$ will have to be isomorphic to the rational elliptic
surface $J$ coming from the $E_{8}$ del Pezzo corresponding to $W$. In
this setting the stable degeneration consists of two copies of $J$
which are identified along $E$ sitting inside each of them only the
gluing map is not the identity on $E$ but rather the automorphism
$t_{\alpha} : E \rightarrow E$. We will denote the resulting normal
crossing surface by $J\coprod_{E, t_{\alpha}} J$. By construction
$J\coprod_{E, t_{\alpha}} J$ has a fixed point free involution which
interchanges the two copies of $J$ and acts as $t_{\alpha}$ on the
common curve $E$. 

\subsec{The quotient model}

If one thinks heuristically about the CHL string as a Heterotic string
with involution one  expects that the duality should produce an F-theory
compactified on a K3 with involution. Both on-the-nose
correspondence and its stable limit have this property but
nevertheless the duality they produce is not completely
satisfactory. One way to see that is to observe that if the pair
$(S,e)$ is to correspond to the CHL string, then the corresponding
F-theory will have to be compactified on the variety $S/\langle e
\rangle$. This however immediately runs into a problem since $S/\langle e
\rangle$ is an Enriques surface and hence cannot be Ricci flat.
This seeming contradiction is resolved if we go back and carefully
compare the action of the Enriques involution $e$ and the CHL
involution $i$ on the Narain lattice ${\bf Nar} \subset \Lambda$.
Recall that (up to a rescaling by $2$ on the second factor)
we identified the CHL lattice ${\bf Nar}^{\langle i \rangle} = {\bf  
U}\oplus
{\bf U} \oplus {\bf E}_{8}(2)$ with the transcendental lattice  
${\bf U}\oplus
{\bf U}(2) \oplus {\bf E}_{8}(2)$ of $S$ which in turn is identified
with the lattice of {\it anti-invariants} of $e$ acting on
$\Lambda$. In particular, we see that the action of $i$ on the Narain
lattice matches\foot{After embedding ${\bf Nar} \subset
\Lambda$ as $(u, v, \xi', \xi'') \mapsto (u, v, -v, \xi', \xi'')$.}
the action of the involution $-e$ (rather than that of $e$). Therefore
if we want to identify the action of the CHL involution on the
F-theory side we need to find a geometric realization of the
involution $-e$, that is we need to find an involution $\sigma : S
\rightarrow S$ which on the Narain part of the cohomology of $S$
induces exactly\foot{Up to a choice of marking.} the involution $-e$.

This can be easily done if we remember that by virtue of the
Heterotic/F-theory duality the surface $S$ comes equipped with an
elliptic fibration with a section. Let ${\bf inv}$ denote the
involution on $S$ which acts as inversion along the elliptic fibers.
We will see that ${\bf inv}$ and $e$ commute as
automorphisms of $S$ and that the composition $\sigma := {\rm
inv}\circ e$ acts as $-e$ on the Narain lattice ${\bf Nar} \subset  
\Lambda$.

The key is understanding the relationship of $\sigma$ and the Enriques
surface $F = S/\langle e \rangle$. Since the elliptic fibration
structure on $S$ induces an elliptic
fibration $a : F \rightarrow {\bf P}^{1}$ on $F$ one can change the
viewpoint and consider both $S$ and $\sigma$ as attributes of $F$ and
its  elliptic fibration structure. For this we will need some
preliminaries on the moduli space of elliptic Enriques surfaces.

It is well known (see \ref\bpv{Barth,  W.; Peters, C.; van de Ven,
A. {\it Compact complex
surfaces.} Ergebnisse
der Mathematik und ihrer Grenzgebiete (3)  4. Springer-Verlag, 1984},
\dolenr, Chapter V or \ref\grhar{P.Griffiths,
J.Harris, {\it Principles of algebraic geometry,} Wiley-Interscience,
New York, 1978.}, Section 4.5) that every Enriques surface admits an
elliptic pencil. Since the general Enriques surface has an infinite
automorphism group (see \ref\bp{W.Barth, C.Peters, {\it Automorphisms
of Enriques surfaces,} Invent. math. 73 (1983), 383-411.}) this
implies that the general Enriques surface admits infinitely many
elliptic structure. However only finitely many\foot{527 to be precise
- see \bp.} of those are
inequivalent modulo automorphisms and so the moduli space of elliptic 
Enriques surfaces is just a finite cover of the moduli of Enriques
surfaces. Every elliptic fibration $a : F \rightarrow {\bf P}^{1}$ on
an Enriques surface $F$ has exactly
two multiple\foot{Recall that a fibration $f : \Sigma \rightarrow B$ of a
smooth variety $\Sigma$ over a curve $B$ is said
to have an $m$-tuple fiber at a point $b \in B$ if the pullback
$f^{*}u$ of a local coordinate $u$ at $b$ vanishes to order $m$ at
{\it every} point of $f^{-1}(b)$.} fibers
(see e.g. \grhar, Section 4.5). In particular $a$ cannot have a
section but it is known that $a$ admits a 2-section of arithmetic
genus $0$ or $1$. The relative Jacobian $a^{0} : J^{0}(F/{\bf P}^{1})
\rightarrow {\bf P}^{1}$ of $a$ is an elliptic surface with a section
and twelve $I_{1}$-fibers, i.e. a rational elliptic surface
\dolenr. A remarkable construction of Kodaira - the so called
logarithmic transform\foot{The logarithmic transform is a general  
procedure which produces
multiple fibers in elliptic fibrations. Here is a brief description of
how the logarithmic transform works (see \grhar\ for more
details). Start
with an elliptic surface $f :
\Sigma \rightarrow B$, a point $b \in B$ and a non-trivial
$m$-torsion point $\alpha_{b} \in \Sigma_{b}$. The existence of  
$\alpha_{b}$
implies that $\Sigma_{b}$ has a non-trivial fundamental group,
i.e. that $\Sigma_{b}$ is of type $I_{k}$, $k \geq 0$. For simplicity
we assume that $k = 0$, i.e. that $\Sigma_{b}$ is a smooth elliptic
curve. If $b \in \Delta \subset B$ is a small analytic disk, then
$\Sigma_{|\Delta} \rightarrow \Delta$ is a topologically trivial
elliptic fibration since $\Delta$ is simply connected. In particular
the cover of $m$-torsion points splits over $\Delta$ and so there is a
unique analytic section $\alpha : \Delta \rightarrow \Sigma_{|\Delta}$
of $m$-torsion points passing trough $\alpha_{b}$. Let $p_{m} :  
\Delta_{m}
\rightarrow \Delta$ be the degree $m$ cyclic cover of $\Delta$
branched at $b$. The multiplication by an $m$-th root of unity on
$\Delta_{m}$ lifts to an automorphism $\phi :
p_{m}^{*}(\Sigma_{|\Delta}) \rightarrow p_{m}^{*}(\Sigma_{|\Delta})$ (of
order $m$). The translation $t$ by the (pull-back) of the section
$\alpha$ is also an automorphism of order $m$ of
$p_{m}^{*}(\Sigma_{|\Delta})$ and the quotient
$(p_{m}^{*}(\Sigma_{|\Delta}))/\langle \phi\circ t \rangle$ is an  
analytic
surface isomorphic to $\Sigma_{|\Delta}$ over the punctured disk
$\Delta - \{ b \}$ and having an $m$-tuple fiber at $b$. By gluing
back this quotient to $\Sigma - \Sigma_{b}$ one obtains a new complex
surface ${\bf Log}(\Sigma, b, \alpha_{b})$ which is analytically
isomorphic to $\Sigma$ over $B - \{ b \}$ and has a multiple fiber  
at $b$.}
 - allows one to
reconstruct $F$ from $J^{0}(F/{\bf P}^{1})$ and a little bit of extra
data. In this case the extra data needed is a choice of two smooth  
elliptic
fibers of the rational elliptic surface $J^{0}(F/{\bf P}^{1})$ and a
choice of a point of order two on each of them. To summarize - the
moduli space of elliptic Enriques surfaces can be identified with the
moduli space of triples $(J, (E_{1},\alpha_{1}),  (E_{2},\alpha_{2}))$
where $J$ is a rational elliptic surface, $E_{1}$ and $E_{2}$ are two
anticanonical curves on $J$ with marked origin and and $\alpha_{i} \in
E_{i}[2]$ are points of order two. Since every rational elliptic
surface is the blow-up of ${\bf P}^{2}$ in the nine intersection
points of two cubic curves, we obtain an equivalent description of the
moduli space as the space parameterizing pairs of plane cubics with a
choice of a point of order two on each of them. The rational elliptic
surface $J$ corresponding to $F$ is exactly the one appearing in the
stable degeneration of $S$ and pair $(E,\alpha)$ appearing in the CHL
data determining $F$ is isomorphic to one of the two pairs $(E_{1},
\alpha_{1})$, $(E_{2}, \alpha_{2})$. Thus the moduli of CHL data is
actually isomorphic to the moduli space of elliptic Enriques surfaces
with a marked double fiber.

With this in mind we can now go back and reexamine the involution  
$\sigma :
S \rightarrow S$. The data of $F$ and its elliptic fibration $a : F
\rightarrow {\bf P}^{1}$  determines two points on the base ${\bf
P}^{1}$ - namely the points where the double fibers of $a$ sit. For $B
\cong {\bf P}^{1} \rightarrow {\bf P}^{1}$ - the double cover
branched at those two points we can form the fiber product
$F\times_{{\bf P}^{1}} B$ which is not normal along the curves over
the two marked points. The normalization $(F\times_{{\bf P}^{1}}
B)^{\nu}$ is a smooth non-ramified double cover of $F$ which is
easily seen to be irreducible and so coincides with $S$. If $x \in
{\bf P}^{1}$ is a point different from the two marked points the
preimage of $x$ in $B$ consists of two points $b'$ and $b''$ and the
preimage of $x$ in $S$ consists of two copies $E'$ and $E''$ of the
fiber $F_{x}$. The Enriques involution $e$ acts by interchanging $E'$
and $E''$ and the inversion acts on $E'$ and $E''$ separately. In
particular $e$ and ${\bf inv}$ commute outside of the preimages of the
double fibers and since these are global automorphisms they must
commute everywhere. This implies that $\sigma = e\circ {\bf inv} =
{\bf inv}\circ e$ is an involution on $S$ and by the description  
just given
we see that $\sigma$ acts without fixed points outside of the
preimages of the two double fibers in $F$. Furthermore the action of
$\sigma$ leaves each of the two preimages of the double fibers and has
for fixed points on each of them. To understand what the quotient
$X := S/\langle \sigma \rangle$ looks like note that since both $e$
and ${\bf inv}$ act as multiplication by $-1$ on the holomorphic
symplectic form $\Omega$ on $S$, the composition $\sigma$ will leave
$\Omega$ invariant and so $X$ will be a K3 surface with 8
singularities of type $A_{1}$. Moreover since the elliptic fibration
on $S$ has a section one knows that the quotient $S/\langle {\bf inv}
\rangle$ is isomorphic to the Hirzebruch surface ${\bf F}_{4}$. This
implies that the sublattice of invariants $H^{2}(S,{\bf Z})^{\langle
{\bf inv} \rangle}$ can be identified with the pull-back of the second
cohomology of ${\bf F}_{4}$ and is hence coincides with the sublattice
${\bf U} \subset H^{2}(S,{\bf Z})$ spanned by the elliptic fiber and
the section. In particular ${\bf inv}$ acts as multiplication by $-1$
on the transcendental lattice of $S$ and therefore $\sigma = e\circ
{\bf inv}$ acts as $-e$ on the Narain part of the cohomology of $S$.

As explained in the beginning of the subsection this is exactly the
condition that an involution on $S$ should satisfy in order to match
the CHL involution. Thus the F-theory corresponding to the CHL string
is naturally compactified on the K3 surface $X = S/\langle \sigma
\rangle$. The CHL lattice ${\bf Nar}^{\langle i \rangle} = {\bf
U}\oplus  {\bf U}\oplus {\bf E}_{8}$
appears in this context as the intersection of the invariants of the
$\sigma$ action on $H^{2}(S,{\bf Z})$ and the transcendental lattice
of $S$. In other words the CHL latice becomes exactly the (pull-back
to $S$) of the transcendental lattice of $X$. Alternatively we may
blow up the eight $A_{1}$ singularities of $X$ to obtain a smooth K3
surface $\hat{X}$ whose Picard lattice contains the even non-degenerate
lattice $N$ of signature $(1,9)$. Explicitly the elliptic fibration on
$S$ descends to an elliptic fibration on $p: X \rightarrow {\bf
P}^{1}$ with two special fibers which are double ${\bf P}^{1}$'s each
carrying four of the eight singular points. The elliptic fibration $p$
does not have a section but has a double section and so the
elliptically fibered K3 surface $\hat{p} : \hat{X} \rightarrow {\bf
P}^{1}$ has two fibers of type $I^{*}_{0}$ (=$\hat{D}_{4}$) and a
double section intersecting the {\it double} component of each
$I^{*}_{0}$ fiber at one point. Therefore the lattice $N \subset {\rm
Pic}(\hat{X})$ is generated by the ten components of the two
$I^{*}_{0}$ fibers and the double section of $\hat{p}$. In this way
the moduli space of F-theory compactifications dual to the CHL string
becomes naturally isomorphic to the moduli space of all $N$-polarized
K3 surfaces or equivalently to the moduli space of all elliptic K3
surfaces with two $I^{*}_{0}$ fibers and a double section which
recovers Witten's description from \toroidal.  

To justify the last identification we have to exhibit a way of
reconstructing $S$ from the surface $X$.  To do that consider the
resolved model $\hat{p} : \hat{X} \rightarrow {\bf P}^{1}$ and let
$p_{1}, p_{2} \in {\bf P}^{1}$ be the two points over which we have
the two $I^{*}_{0}$ fibers. The fibers over the $p_{i}$'s can be
written as
\eqn\fibers{
\hat{X}_{p_{i}} = 2f_{i} + e^{1}_{i} + e^{2}_{i} + e^{3}_{i} +
e^{4}_{i}, \; i = 1, 2
}
where $f_{1}, f_{2}$ and  $e^{j}_{i}$, $i= 1, 2$, $j = 1,2,3,4$ are
smooth $-2$ curves and the $e^{j}_{i}$'s contract to the eight
singular points of $X$. Consider the line bundle $L = {\cal
O}_{\hat{X}}(\sum_{j =  1}^{4} (e^{j}_{1} + e^{j}_{2}))$ on
$\hat{X}$. By \fibers\ we have $L = {\cal O}_{\hat{X}}(\hat{X}_{p_{1}}
+ \hat{X}_{p_{2}} - 2f_{1} - 2f_{2}) = \hat{p}^{*}{\cal O}_{{\bf
P}^{1}}(2)\otimes {\cal O}_{\hat{X}}(- 2f_{1} - 2f_{2})$ and so $L$
has a unique square root equal to $\hat{p}^{*}{\cal O}_{{\bf
P}^{1}}(1)\otimes {\cal O}_{\hat{X}}(- f_{1} - f_{2})$. Due to this
there exists a unique double cover $\hat{S} \rightarrow \hat{X}$ which
is simply branched exactly at the eight curves $e^{j}_{i}$. The
ramification divisor of the cover $\hat{S} \rightarrow \hat{X}$
consists of eight smooth rational curves $\hat{e}^{j}_{i}$ mapping
isomorphically to the branch curves $e^{j}_{i}$. Since the
$\hat{e}^{j}_{i}$'s are part of the ramification divisor we have
$(2\hat{e}^{j}_{i})^{2} = 2 (e^{j}_{i})^{2} = -4$ and hence
$(\hat{e}^{j}_{i})^{2} = -1$ for all $i, j$. Therefore we can contract
all of the $\hat{e}^{j}_{i}$'s to obtain a smooth K3 surface $S$.

\bigskip

\noindent
{\bf Remark} The Enriques surface $F$ played an auxiliary role in the
duality construction above and will not appear in our considerations
from now on. It is worth mentioning however that by replacing the
Enriques involution $e$ by the involution $\sigma$ one introduces a
finite ambiguity. More precisely, the moduli space of pairs
$(S,e)$ is a finite cover of the moduli space of pairs
$(S,\sigma)$. Indeed - to reconstruct a pair $(S,e)$ from a pair
$(S,\sigma)$ one needs also to specify points of order two along the
two $\sigma$-invariant fibers of the Weierstrass elliptic fibration on
$S$.

\subsec{The $\Gamma_{0}(2)$-model}

As we have just seen the F-theory dual to the CHL string lives
naturally on an elliptic K3 surface $X$ of a special kind. The
description we have obtained however is still not completely
satisfactory since as we saw the elliptic fibration structure $p : X
\rightarrow {\bf P}^{1}$ that $X$ inherits from the duality with the
CHL string does not admit a section but only a double section. This
causes a problem since in view of the duality with the IIB string the
F-theory moduli should all be encoded into data living entirely on the
base of the elliptic fibration, i.e. the F-theory should be
compactified on a Weierstrass elliptic fibration. One can try to
remedy the lack of section for $p : X \rightarrow {\bf P}^{1}$ by
replacing $X$ by its (basic) relative Jacobian fibration\foot{Which
turns out to be a K3 surface as well.} $J^{0}(X/{\bf
P}^{1}) \rightarrow {\bf P}^{1}$ which always has a Weierstrass form
and so is encoded into geometric data on ${\bf P}^{1}$ only. This is
not a big sacrifice to make since $X$ is always algebraic and so
according to Kodaira's (see \ref\kodaira{K. Kodaira, {\it Compact
analytic surfaces III,} Ann. of Math.}) and Ogg-Shafarevich's
(see \dolenr\ , Section V.3) theory the surface $X$ is determined
by the surface $J^{0}(X/{\bf P}^{1})$ up to finite data. On the other
hand if the deformation theory of the surface $X$ is governed by a
Weierstrass eqution only we should be seen an 18 dimensional moduli
space rather than a 10 dimensional one since there is no obstruction
to deforming a K3 surface in Weierstrass form away from the locus of
elliptic fibrations having two $I^{*}_{0}$ fibers. Therefore it is
necessary to find a mechanism that is intrinsic to F-theory that will
allow the confinement of the moduli of the elliptic fibration to the
subfamily we get trough the duality with the CHL string.

One instance in which this will necessarily happen is if there is an 
additional field in F-theory whose deformations are obstructed in the 
full moduli of elliptic K3 surfaces. The only field which is
intrinsically present in the  F-theory and which has been set to equal
zero in order for the full  F-theory/Heterotic duality to work is the
skew-symmetric B-field on the base of the F-theory
compactification. So it is reasonable to expect that in a consistent
realization of the F-theory vacua on $X$ one will have to describe not
only the dilaton-axion fields but also a B-field which constrains the
deformations of the compactification. To that end we examine how the
presence of a non-trivial B-field is encoded in the geometry of an
F-theory compactification.

Let $a: CY \rightarrow S$ be an elliptically
fibered Calabi-Yau manifold on which F-theory is compactified. Fix a
point $o \in S^{o} := S\setminus{\rm Discriminant}$ in the  
complement of the
discriminant of $a$. Let $E = a^{-1}(o)$ be the fiber of $a$ over
$o$ and let ${\bf mon} : \pi_{1}(S^{o}, o)
\rightarrow SL(2,{\bf Z}) \subset GL(H^{1}(E,{\bf Z}))$ be the
monodromy representation of the family $X \rightarrow S^{o}$.
The data describing a B-field in
the F-theory consists of a complex multivalued 2-form $\zeta$ on
$S^{o}$ which satisfies: (i)  $\zeta$ becomes
single valued modulo the translation action of the lattice
$H^{1}(E,{\bf Z})$; (ii) the pair $(\zeta, \bar{\zeta})$ transforms
according to the monodromy representation ${\bf mon}$.  In
coordinate-free terms $\zeta$ should be viewed as a multivalued 2-form 
on $S^{o}$ with coefficients in the
complex vector space $H^{0,1}(E) = H^{1}(E,{\cal O}_{E})$ for which: 
(i) $\zeta$ is single valued modulo the translation action of the  
lattice
$H^{1}(E,{\bf Z}) \subset H^{1}(E,{\cal O}_{E})$; (ii) $(\zeta,
\bar{\zeta}) \in H^{0,1}(E)\oplus H^{1,0}(E) = H^{1}(E,{\bf C})$ is 
invariant under the action of ${\bf mon}(\pi_{1}(S^{o},o))$. Since
only the cohomology class of these 2-forms on $S$
is visible by the physics, after fixing a basis in $H^{2}(S,{\bf C})$ 
we can view the components of $\zeta$ as real multisections
in $R^{1}a_{*}{\bf C}\otimes H^{2}(S,{\bf C})$ which are horizontal
(w.r.t. the Gauss-Manin connection) and become single valued modulo
the action of $R^{1}a_{*}{\bf Z} \subset R^{1}a_{*}{\bf C}$. This
implies in particular that  when projected to $R^{1}a_{*}{\cal
O}_{X}/R^{1}a_{*}{\bf Z} \subset CY$ each component of $\zeta$ becomes a
holomorphic (single valued) section of the elliptic fibration $a : CY
\rightarrow S$ which is horizontal\foot{Or in other words is a locally
constant normal function in the sense of \ref\gr{P.Griffiths, {\it
Infinitesimal variations of Hodge structures (III): determinantal
varieties and the infinitesimal invariant of normal functions.}
Comp. Math. 50 (1983) 267-324.}.} for the Gauss-Manin connection on
the family of elliptic curves $CY_{|S^{o}} \rightarrow
S^{o}$. Conversely, any such section of $a$ will determine the
component of a B-field.

The situation simplifies somewhat if we are in the case where
$h^{2}(S,{\bf C}) = 1$. In this case the B-field $\zeta$ itself is
realized as a holomorphic section of $a : CY \rightarrow S$ which is
Gauss-Manin horizontal. If the Calabi-Yau manifold $CY$ is general in
the sense that ${\bf mon}(\pi_{1}(S,o)) = SL(2,{\bf Z})$, then the
only such section will be the zero section of the elliptic
fibration. So generically the B-field will be zero. Moreover it is not
hard to see that the only horizontal sections for the Gauss-Manin
connection will have to be sections representing torsion points on the
elliptic fiber. Thus the only situations when one can turn-on the
B-field in the F-theory compactified on $CY$ is when the monodromy
group ${\bf mon}(\pi_{1}(S,o))$ preserves some vector in ${1 \over
n}H^{1}(E,{\bf Z}) \subset H^{1}(E,{\bf Q})$, i.e. when ${\bf
mon}(\pi_{1}(S,o))$ is contained in some congruence subgroup of level
$n$.

Going back to the compactification on the special K3 surface $X$ one
can observe that the inherited elliptic fibration $p : X \rightarrow
{\bf P}^{1}$ will have general monodromy and so one cannot turn on the
B-field neither for the fibration on $X$ nor for the fibration on
$J^{0}(X/{\bf P}^{1})$. Therefore, in order to obtain a consistent
description for the F-theory vacua on $X$ we are forced to look for  
a {\it
different} elliptic structure on $X$ which has monodromy contained in
a congruence subgroup. We find a natural elliptic structure $\pi : X
\rightarrow {\bf P}^{1}$ like that
which turns out to have monodromy contained in the congruence subgroup
$\Gamma_{0}(2)$. This means that $\pi$ admits a section consisting of
2-torsion points or equivalently that the F-theory compactified on $X$
with the elliptic fibration $\pi$ admits a non-trivial B-field
corresponding to $1/2$ the Fubini-Studi form on ${\bf P}^{1}$. As a 
consistency check we study the moduli space of
deformations of this elliptic fibration that preserve the B-field and
show that it is isogeneous to the moduli space of $X$ considered as a
$N$-polarized K3 surface. Further evidence for the relevance of the
$\Gamma_{0}(2)$ fibration was provided in Section~3 where we describe an
effective mechanism  allowing one to read the spectrum of gauge
groups for the CHL string in terms of the elliptic fibration $\pi$.

The construction of the fibration $\pi : X \rightarrow {\bf P}^{1}$ or
equivalently its lift $\hat{\pi} : \hat{X} \rightarrow {\bf P}^{1}$ to
the resolution of $X$ is based on a specific projective model of $X$
which we explore next.

We have already seen one explicit model of $S$ and $X$ in
Section~2. There $S$ was given in its Weierstrass form \fkkk\ and $X$
was described as the quotient of $S$ by the involution $\sigma(x,y:z)
= (x,-y;1/z)$. The elliptic fibration on $S$ descends through this
construction to the inherited elliptic fibration $p : X \rightarrow
{\bf P}^{1}$.  To obtain the $\Gamma_{0}(2)$ elliptic fibration $\pi : X
\rightarrow {\bf P}^{1}$ we will construct $X$ in a different way -
namely as a double cover of the Hirzebruch surface $Q := {\bf F}_{0} =
{\bf P}^{1}\times {\bf P}^{1}$. It is well known that all Enriques
surfaces can be realized as double covers of $Q$ \dolenr\. This
suggests that the surfaces $X$ can also be realized as double covers
of quadrics which is indeed the case.

In order for a double cover of $Q$ to be a K3 surface it has to be
branched along a curve of type $(4,4)$, i.e. along the divisor of a
section in the line bundle ${\cal O}_{Q}(4,4)$. Every such cover will
have two natural elliptic fibrations corresponding to the two rulings
of the quadric $Q$, that is to the projections $q_{1} : Q \rightarrow
{\bf P}^{1}$  and $q_{2} : Q \rightarrow {\bf P}^{1}$ to the two factors
of $Q$. If we want to have two $I^{*}_{0}$ fibers in say the fibration
corresponding to $q_{1}$ we are necessarily forced to consider only  
double
covers of $Q$ branched along curves of the form $r_{1} \cup r_{2} \cup
C$ where the $r_{i}$'s are two fibers of $q_{1}$ and $C$ belongs to the
linear system ${\cal O}_{Q}(2,4)$. The number of moduli of such K3's
is easily seen to be 10 since we have $\dim {\bf P}(H^{0}(Q,  
q_{1}^{*}{\cal
O}_{{\bf P}^{1}}(2))) = \dim {\bf P}(H^{0}({\bf P}^{1}, {\cal
O}_{{\bf P}^{1}}(2))) = 2$ parameters in the $r_{i}$'s and $\dim
{\bf P}(H^{0}(Q,{\cal O}_{Q}(2,4))) = \dim
{\bf P}(H^{0}({\bf P}^{1}, {\cal
O}_{{\bf P}^{1}}(2))\otimes H^{0}({\bf P}^{1}, {\cal
O}_{{\bf P}^{1}}(4))) = 3\cdot 5 - 1 = 14$ parameters in the curves
$C$ and we also have the 6 dimensional group of automorphisms of $Q$. 
Due to this the moduli space of $N$-polarized K3 surfaces $X$ is
isomorphic to the moduli space of double covers of $Q$ branched along
curves of the form $r_{1} + r_{2} + C$ and so each $X$ has a second
elliptic fibration $\pi : X \rightarrow {\bf P}^{1}$ corresponding to
the projection $q_{2} : Q \rightarrow {\bf P}^{1}$.

The elliptic fibration $\pi$ has two sections $f_{1}$ and $f_{2}$,
each corresponding to the double component of the two $I^{*}_{0}$
fibers in the elliptic fibration $p$. Each of the $f_{i}$'s passes
trough four of the $A_{1}$ singularities of $X$ as seen on
Figure~4. On the resolved surface $\hat{X}$ then one sees that the
elliptic fibration $\hat{\pi} : \hat{X} \rightarrow {\bf P}^{1}$ has
eight fibers of type $I_{2}$ and two sections arranged in such a way
that at each $I_{2}$ fiber the two sections pass trough two different
components of the fiber. It is not hard now to obtain a representation
of the moduli space of $X$'s as a Weirestrass family - namely we can
declare the section $\hat{f}_{1}$ to be the
zero section and contract the components in the $I_{2}$ fibers not
meeting it. The resulting surface $Y$ will be an elliptic K3 surface with
two sections and  eight $A_{1}$ singularities sitting on one of them.
Moreover if we examine what this operation does to the realization of
$X$ as a double cover of the quadric $Q$ we see that the passage from
$X$ to $Y$ just amounts to performing $4$ elementary modifications of
$Q$ at the four points of intersection of $r_{1}$ and $C$. In other
words on $Q$ we will have to blow up the four points $\{ x_{1}, x_{2},
x_{3}, x_{4} \} = C \cap r_{1}$ and then contract the $q_{2}$-fibers
$q_{2}^{-1}(q_{2}(x_{1})), q_{2}^{-1}(q_{2}(x_{2})),
q_{2}^{-1}(q_{2}(x_{3})), q_{2}^{-1}(q_{2}(x_{4}))$ passing trough
these points. Since each elementary modification like that increases
the index of the Hirzebruch surface by one at the end of the day we
will get a realization of $Y$ as a double cover of the Hirzebruch
surface ${\bf F}_{4}$. In particular the K3 surface $Y$ is in a
Weierstrass form so that the section $f_{1}$ sits over the section
at infinity\foot{i.e. the unique section of self intersection
$-4$.}  in ${\bf F}_{4}$ and is thus interpretted as the zero
section of the elliptic fibration $w : Y \rightarrow {\bf P}^{1}$. The
curve of points of order two for $w$ is then naturally identified with
the ramification divisor of the double cover $Y \rightarrow {\bf
F}_{1}$ and consequently coincides with the (proper transform) of the
curve $f_{1}\cup f_{2}\cup C$. Due to this we conclude that both $w : Y
\rightarrow {\bf P}^{1}$ and $\pi : X \rightarrow {\bf P}^{1}$ have 
monodromy contained in $\Gamma_{0}(2)$ and so
the second section $f_{2}$ of $w$ should be viewed as the B-field for
the F-theory compactification\foot{Note that both $X$ and $Y$ have a
common resolution $\hat{X}$ on $X$.}.

Finally we describe the deformations of the Weierstrass fibration $w : Y
\rightarrow {\bf P}^{1}$ in which the second section $f_{2}$
survives. For that one just needs to observe that all such
deformations are coming as double covers of ${\bf F}_{4}$ branched
along a curve of the form $e_{\infty} + e_{0} + C$ where $e_{\infty}$
is the infinity section of ${\bf F}_{4}$, $e_{0}$ is a section in
${\cal O}_{{\bf P}^{1}}(4) \subset {\bf P}({\cal O}_{{\bf
P}^{1}}\oplus {\cal O}_{{\bf P}^{1}}(4))$ and $C$ is a curve in the
linear system $|2e_{\infty} + 8f|$ ($f$ is the fiber of ${\bf F}_{4}
\rightarrow {\bf P}^{1}$. The section $e_{0}$ is then given by a
degree four polynomial $a_{4}(z)$ on ${\bf P}^{1}$ and so the Weierstrass
equation of $Y$ must be of the form \model\ we used in Section~3.

\subsec{Matching of collision patterns}

The representation of the surface $X$ as double covers of a quadric is
very useful not only for finding the $\Gamma_{0}(2)$ elliptic fibration on
$X$ but also for analyzing the gauge groups appearing after several
singular fibers of $\pi$ collide in the presence of a non-trivial
B-field. Here we show how to carry such an analysis in several model
cases. This information was used extensively in Section~3 where we
described the gauge group spectrum of the F-theory with fluxes on $X$.

The basic idea which makes the matching of singularities possible is
the observation that the singular fibers of either the inherited or the
$\Gamma_{0}(2)$ elliptic fibrations on $X$ will correspond to certain
singularities of the branch curve $r_{1}\cup r_{2} \cup C$. Thus we
may describe the Kodaira fibers (in either fibration) on the minimal
resolution of $X$ by just resolving the singularities on the branch
curve and keeping track of what part of the exceptional locus will
have to enter the resolved branch divisor.

\bigskip

\noindent
{\it $SU(n)$ gauge groups:} As usual we are interested only in
non-abelian gauge groups so we are in the case when $n \geq
2$. According to our recipe the $SU(n)$ gauge groups appear when $n +
1$ fibers of type $I_{1}$ in the inherited elliptic fibration collide
with each other and the collision occurs away from the two $I^{*}_{0}$
fibers. In terms of the branch curve this just means that the curve
$C$ acquires a higher order of contact with a ruling $f$ of $p_{2} : Q
\rightarrow {\bf P}^{1}$ which is different from $r_{1}$ and
$r_{2}$. Since we want our gauge groups to be localized at a single
fiber in {\bf both} fibration we will further assume that the high
order of contact of $C$ and $f$ occurs at a single point of $f$.

For example a $I_{1}$ fiber in the inherited elliptic fibration
corresponds to $C$ being tangent to $f$ at one point $x$ and (since
$f\cdot C = 4$) necessarily intersecting it at two more points $x_{1},
x_{2}$. The
corresponding $I_{1}$ fiber of $\pi : X \rightarrow {\bf P}^{1}$ is
then just the double cover of $f$ branched at the divisor $C\cdot f =
2x + x_{1} + x_{2}$.

To get the group $SU(2)$ we need to look at the case when $C\cdot f$
contains a point $x$ with multiplicity $2$ so that $C$ has a node at
$x$. Figure~5a shows how the minimal resolution of $X$ is obtained by
resolving the singularities of the branch divisor and how that leads
to a $I_{2}$ Kodaira fiber in the inherited elliptic fibration. The
branch and ramification divisors are represented by the thick solid lines and 
$f$ and $\phi$ the fibers of the inherited and $\Gamma_{0}(2)$
fibrations respectively as well as their projection to $Q$. The same
singularity viewed from the point of view of the
$\Gamma_{0}(2)$-fibration is shown on Figure~5b.

\vskip .2in
\let\picnaturalsize=N
\def\picsize{5in}
\def\picfilename{su2a.eps}
\ifx\nopictures Y\else{\ifx\epsfloaded Y\else\fi
\global\let\epsfloaded=Y
\centerline{\ifx\picnaturalsize N\epsfxsize \picsize\fi
\epsfbox{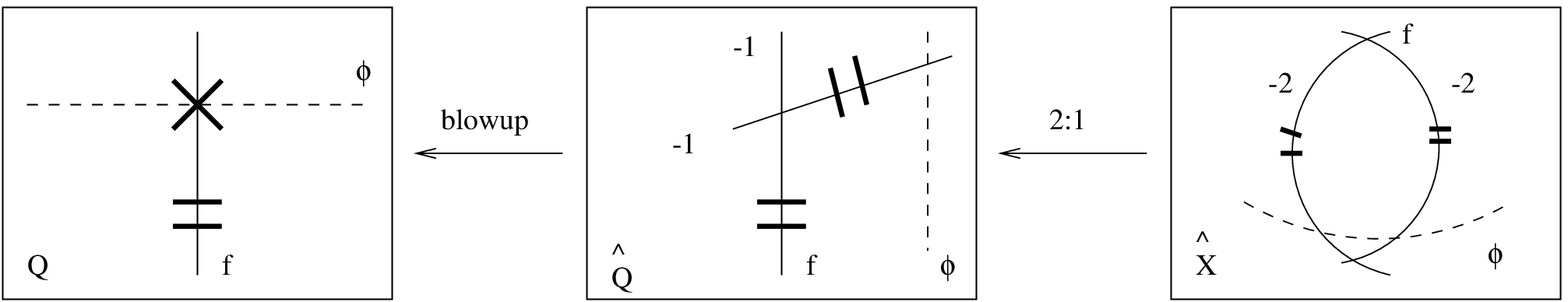}}}\fi
\vskip -.0in

\centerline{Fig.5a: $SU(2)$ in the inherited elliptic fibration}

\vskip .2in
\let\picnaturalsize=N
\def\picsize{5in}
\def\picfilename{su2b.eps}
\ifx\nopictures Y\else{\ifx\epsfloaded Y\else\fi
\global\let\epsfloaded=Y
\centerline{\ifx\picnaturalsize N\epsfxsize \picsize\fi
\epsfbox{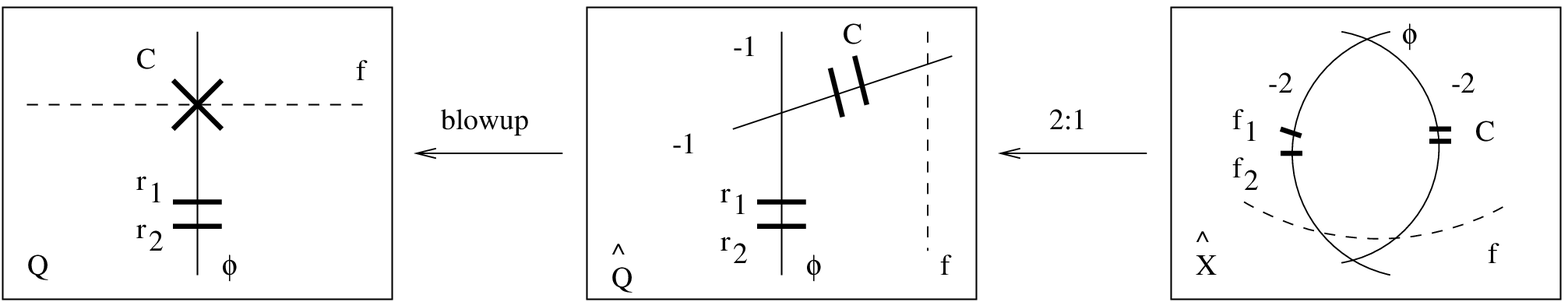}}}\fi
\vskip -.0in

\centerline{Fig.5b: $SU(2)$ in the $\Gamma_{0}(2)$ elliptic fibration}

\bigskip

In exactly the same way a $SU(3)$ gauge group will appear when $C$
acquires a cusp at some point $x \in f$. 

\vskip .2in
\let\picnaturalsize=N
\def\picsize{5in}
\def\picfilename{su3a.eps}
\ifx\nopictures Y\else{\ifx\epsfloaded Y\else\fi
\global\let\epsfloaded=Y
\centerline{\ifx\picnaturalsize N\epsfxsize \picsize\fi
\epsfbox{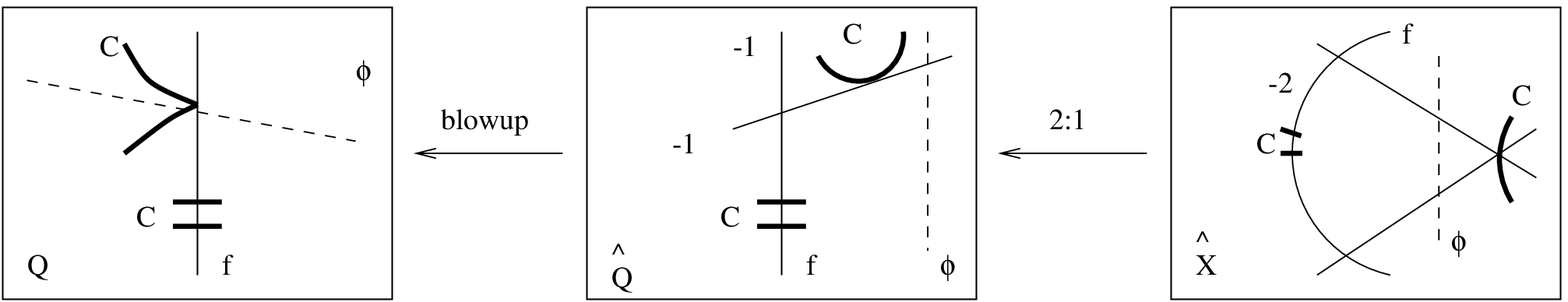}}}\fi
\vskip -.0in

\centerline{Fig.6a: $SU(3)$ in the inherited elliptic fibration}

\vskip .2in
\let\picnaturalsize=N
\def\picsize{5in}
\def\picfilename{su3b.eps}
\ifx\nopictures Y\else{\ifx\epsfloaded Y\else\fi
\global\let\epsfloaded=Y
\centerline{\ifx\picnaturalsize N\epsfxsize \picsize\fi
\epsfbox{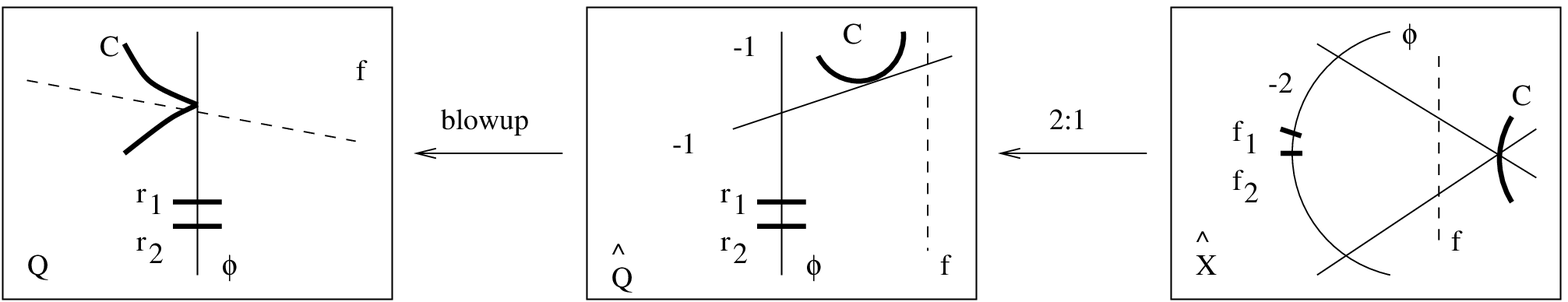}}}\fi
\vskip -.0in

\centerline{Fig.6b: $SU(3)$ in the $\Gamma_{0}(2)$ elliptic fibration}

\bigskip

The groups $SU(n)$ for $n > 3$ are obtained in exactly the same way
after resolution of the appropriate singularities of $C$.

\bigskip

\bigskip

\noindent
{\it $Sp(n)$ gauge groups:} As explained in Section~2 the non-simply
laced gauge groups appear as quotients of Dynkin diagrams of curves on
$S$ by the involution $\sigma$ acting as an exterior automorphisms.
In the inherited elliptic fibration on $S$ this situation corresponds
to collections of $2nI_{1}$ fibers {\it symmetric} with respect to
$\sigma$ colliding at one of the $\sigma$-invariant fibers. In the
inherited elliptic fibration on $X$ this corresponds to $nI_{1}$
fibers colliding on top of one of the $I^{*}_{0}$ fibers and so on $Q$
this is seen as a higher order of contact of the curve $C$ with one of
the rulings entering the branch divisor. For concreteness we will
assume that this ruling is $r_{1}$. 

The first non-abelian case of such collision will occur when the curve
$C$ becomes tangent to $r_{1}$ at one point. The necessary blowups and
the resulting Kodaira fibers in both elliptic fibrations can be seen
on Figures~7a, 7b.

\vskip .2in
\let\picnaturalsize=N
\def\picsize{5in}
\def\picfilename{sp1a.eps}
\ifx\nopictures Y\else{\ifx\epsfloaded Y\else\fi
\global\let\epsfloaded=Y
\centerline{\ifx\picnaturalsize N\epsfxsize \picsize\fi
\epsfbox{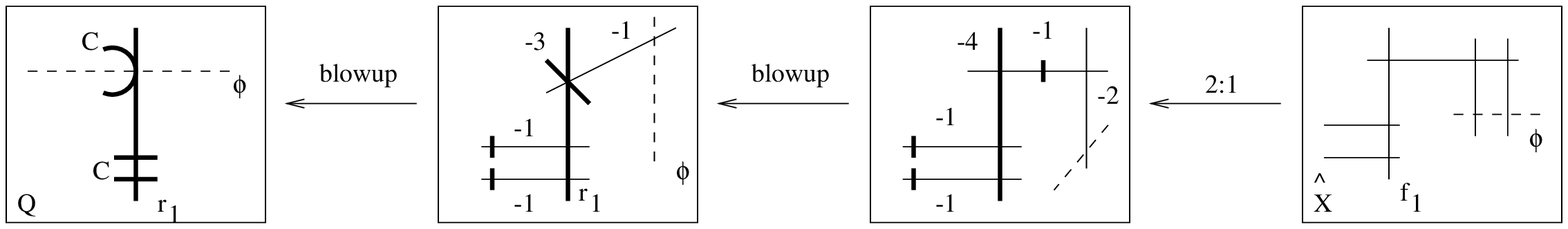}}}\fi
\vskip -.0in

\centerline{Fig.7a: Collision leading to $Sp(1)$ 
(inherited fibration)}

\vskip .2in
\let\picnaturalsize=N
\def\picsize{5in}
\def\picfilename{sp1b.eps}
\ifx\nopictures Y\else{\ifx\epsfloaded Y\else\fi
\global\let\epsfloaded=Y
\centerline{\ifx\picnaturalsize N\epsfxsize \picsize\fi
\epsfbox{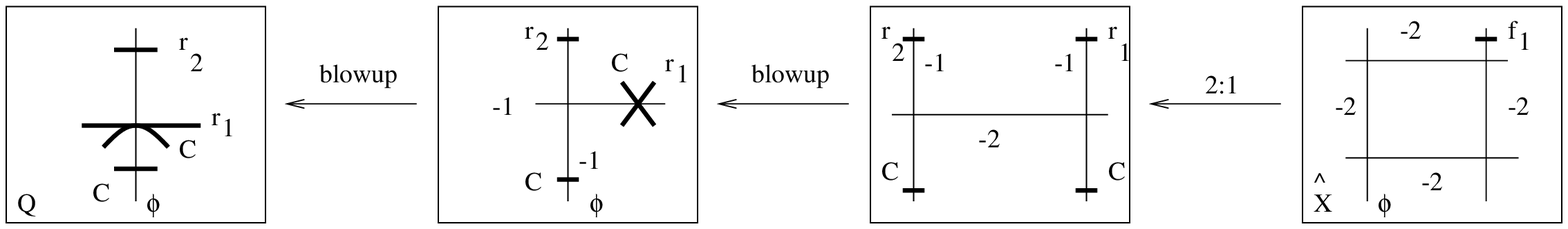}}}\fi
\vskip -.0in

\centerline{Fig.7b: Collision leading to $Sp(1)$ ($\Gamma_{0}(2)$ 
fibration)}

\bigskip

As one can see from these blowups the Kodaira fibers corresponding to
this simplest collision are of type $\hat{D}_{5}$ in the inherited
elliptic fibration and of type $\hat{A}_{3}$ in the $\Gamma_{0}(2)$
elliptic fibration. To see the actual gauge group we have to trace the
behavior of the $-2$ curves comprising these Kodaira fibers on the
double cover $S \rightarrow X$. A convenient way to record what is
happening is to combine both Kodaira fibers described above in one
graph. This graph is shown on Figure~8 where the nodes corresponding
to the inherited fiber are marked as squares and the nodes
corresponding to the $\Gamma_{0}(2)$ fiber are marked as circles. 
The four nodes that are shadowed out are the exceptional
curves corresponding to the four fixed points of $\sigma$ on the
preimage of $r_{1}$ in $S$.

\vskip .2in
\let\picnaturalsize=N
\def\picsize{2in}
\def\picfilename{sp1dyn.eps}
\ifx\nopictures Y\else{\ifx\epsfloaded Y\else\fi
\global\let\epsfloaded=Y
\centerline{\ifx\picnaturalsize N\epsfxsize \picsize\fi
\epsfbox{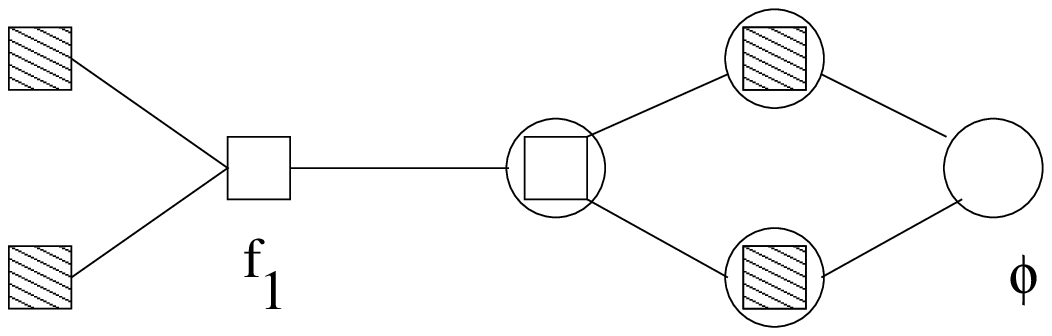}}}\fi
\vskip -.0in

\centerline{Fig.8: Combined Kodaira fibers producing $Sp(1)$}

\bigskip

In particular we see that of the four cycles of the $\hat{A}_{3} (=I_{4})$
fiber in the $\Gamma_{0}(2)$-fibration two are coming from fixed
points for $\sigma$ and hence do not contribute to the gauge symmetry,
one (marked $\phi$) intersects the zero section and also does not
contribute and so there is only a rank one contribution to the gauge
group from the one remaining cycle. Since the remaining cycle
intersects with two of the cycles corresponding to fix points we see
that when pulled back to $S$ it becomes an irreducible rational curve 
invariant under $\sigma$ and thus corresponds indeed to the group
$Sp(1)$.

The collision that produces the fibers on Figure~8 looks like
colliding $2I_{2}$ on the section $f_{1}$ from the viewpoint of the
$\Gamma_{0}(2)$-fibration. Indeed - recall that the $I_{1}$ fibers of
the $\Gamma_{0}(2)$-fibration appear when $C$ becomes tangent to
the fibers of $q_{1} : Q \rightarrow {\bf P}^{1}$ and the $I_{2}$
fibers appear when $C$ intersects $R_{1}$ or $r_{2}$. In particular
the collision of  $2I_{2}$ on the section $f_{1}$ will occur when $C$
becomes tangent to $r_{1}$ which is exactly the situation on
Figure~7b.

To see how this fits in a series we work out the next case of
colliding $2I_{2} + I_{1}$ in the
$\Gamma_{0}(2)$-fibration. The necessary blowups and the corresponding Kodaira
fibers for this collision are shown on Figure~9.

\vskip .2in
\let\picnaturalsize=N
\def\picsize{5in}
\def\picfilename{sp2a.eps}
\ifx\nopictures Y\else{\ifx\epsfloaded Y\else\fi
\global\let\epsfloaded=Y
\centerline{\ifx\picnaturalsize N\epsfxsize \picsize\fi
\epsfbox{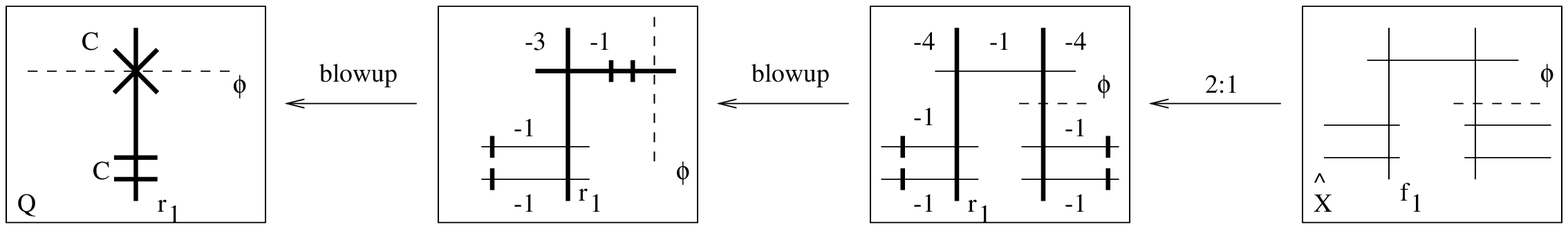}}}\fi
\vskip -.0in

\centerline{Fig.9a: Collision leading to $Sp(2)$ 
(inherited fibration)}

\vskip .2in
\let\picnaturalsize=N
\def\picsize{5in}
\def\picfilename{sp2b.eps}
\ifx\nopictures Y\else{\ifx\epsfloaded Y\else\fi
\global\let\epsfloaded=Y
\centerline{\ifx\picnaturalsize N\epsfxsize \picsize\fi
\epsfbox{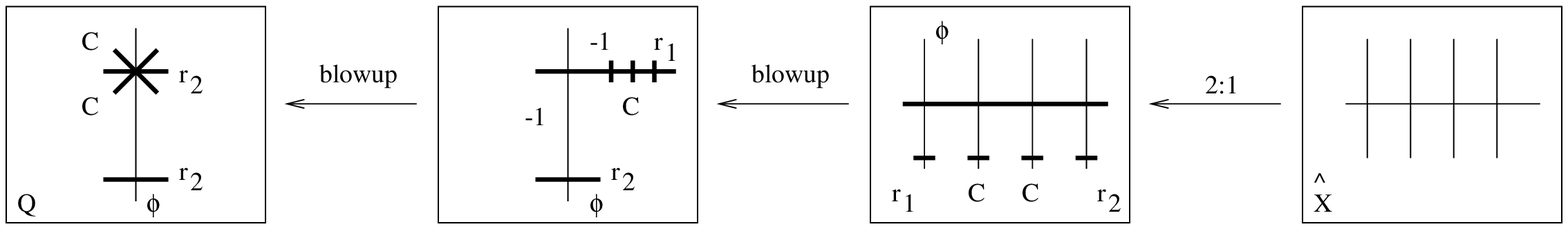}}}\fi
\vskip -.0in

\centerline{Fig.9b: Collision leading to $Sp(2)$ ($\Gamma_{0}(2)$ 
fibration)}

\bigskip

As before the thick black lines denote the curves that are part of the
branch divisor and so we need to blow up enough times and separate
those curves to guarantee that the double cover $\hat{X}$ will be
non-singular. The only new feature appearing here is that the
branch divisor has a singular point with odd multiplicity (a triple
point in this case) and so on the blown-up $Q$ one needs to take not
only the proper transform of $C$ and the $r_{i}$'s as branching but
also one copy of the exceptional divisor in order to make the
corresponding line bundle divisible by 2.

The resulting combined Dynkin diagram is:

\vskip .2in
\let\picnaturalsize=N
\def\picsize{2in}
\def\picfilename{sp2dyn.eps}
\ifx\nopictures Y\else{\ifx\epsfloaded Y\else\fi
\global\let\epsfloaded=Y
\centerline{\ifx\picnaturalsize N\epsfxsize \picsize\fi
\epsfbox{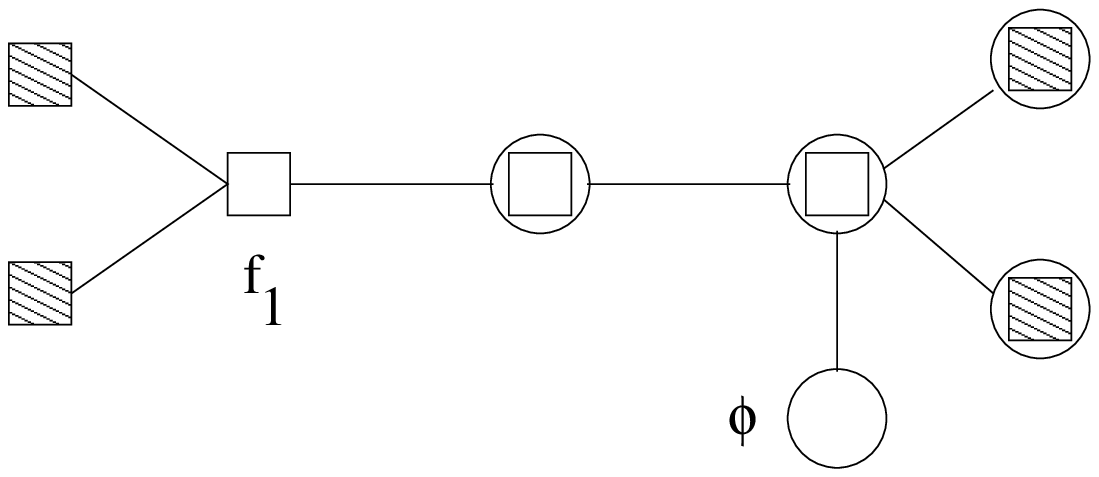}}}\fi
\vskip -.0in

\centerline{Fig.10: Combined Kodaira fibers producing $Sp(2)$}

\bigskip

\noindent
and we see that there are only  two nodes of the $\Gamma_{0}(2)$ fiber that
contribute to the gauge group and that one of them corresponds to an
invariant sphere on $S$. Thus the $\Gamma_{0}(2)$ collision $2I_{2} +
2I_{1}$ with the two $I_{2}$'s coming from $\sigma$ fixed points on
$f_{1}$ corresponds to a gauge group $Sp(2)$.

\bigskip

\noindent
In the same way one can work out the collisions $2I_{2} + nI_{1}$ in
the $\Gamma_{0}(2)$-fibration. The resulting combined Dynkin diagram
is 

\vskip .2in
\let\picnaturalsize=N
\def\picsize{4in}
\def\picfilename{spndyn.eps}
\ifx\nopictures Y\else{\ifx\epsfloaded Y\else\fi
\global\let\epsfloaded=Y
\centerline{\ifx\picnaturalsize N\epsfxsize \picsize\fi
\epsfbox{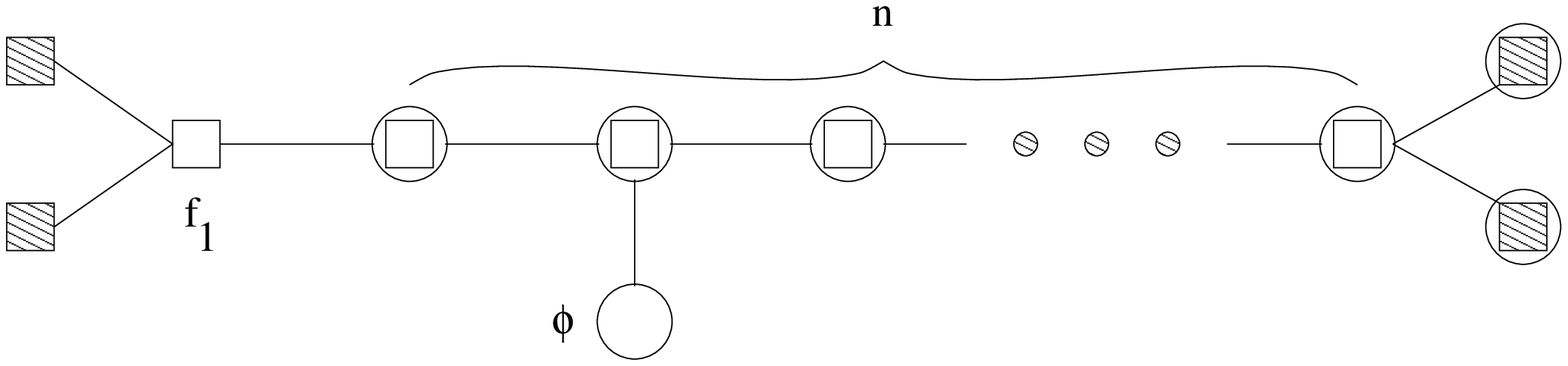}}}\fi
\vskip -.0in

\centerline{Fig.11: Combined Kodaira fibers producing $Sp(n)$}

\bigskip

In summary - for any given collision the gauge group corresponds
exactly to the Dynkin diagram consisting of the cycles which are common to the
two Kodaira fibers and which do not come from fixed points of
$\sigma$. Moreover the cycles intersecting pairs of
exceptional divisors coming from $\sigma$-fixed points correspond to
non-simply laced nodes in the diagram.

\bigskip

\noindent
{\it $SO(2n)$ gauge groups:} Again due to the interpretation of the
CHL string as a Heterotic string with symmetry the level two $SO(2n)$ gauge
groups will appear when on $S$ we have a collection of cycles
intersecting as two copies of a $D_{n}$ Dynkin diagram which are away
from the invariant fibers and are swapped by $\sigma$. In terms of the
inherited elliptic fibration on $X$ this just means that we have a
collision of Kodaira fibers leading to a $\hat{D}_{n}$-fiber
which is away from the two $I^{*}_{0}$ fibers. From the view point of
the quadric $Q$ and the branch curve $C\cup r_{1}\cup r_{2}$ this
collision occurs only when a curve of bidegree $(0,1)$ splits as a
component of the curve $C$ (which has bidegree $(2,4)$). In other
words one of the four points in $C\cap r_{1}$ becomes collinear with
one of the four points in $C\cap r_{2}$. In terms of the
$\Gamma_{0}(2)$ fibration this condition is expressed as a collision
of two $I_{2}$ fibers coming from the two sections $f_{1}$ and
$f_{2}$. Again the gauge groups can be seen as common parts of Kodaira
fibers in the two elliptic fibrations but since the collision takes
place away from the $\sigma$-fixed fibers, there are no shadowed nodes
in the diagram and hence all the nodes are simply laced.

The simplest example is of a collision of $2I_{2}$ fibers in the
$\Gamma_{0}(2)$ fibration which come from different sections. The
relevant blowup pictures are shown on Figure 12.

\vskip .2in
\let\picnaturalsize=N
\def\picsize{5in}
\def\picfilename{so8a.eps}
\ifx\nopictures Y\else{\ifx\epsfloaded Y\else\fi
\global\let\epsfloaded=Y
\centerline{\ifx\picnaturalsize N\epsfxsize \picsize\fi
\epsfbox{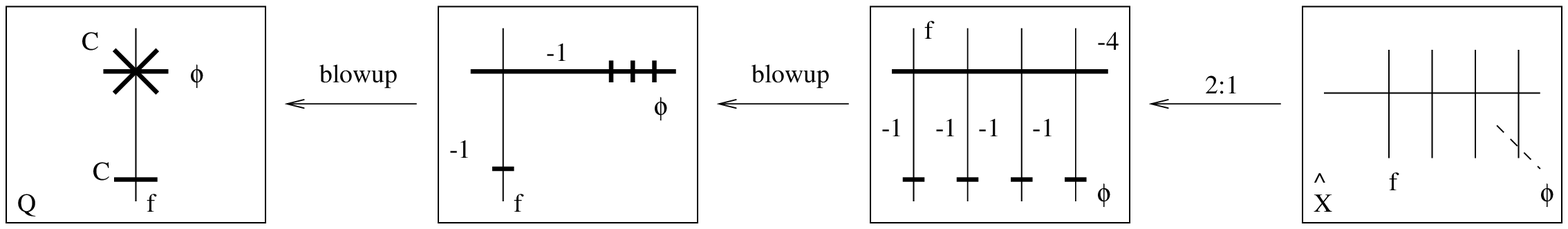}}}\fi
\vskip -.0in

\centerline{Fig.12a: Collision leading to $SO(8)$ 
(inherited fibration)}

\vskip .2in
\let\picnaturalsize=N
\def\picsize{5in}
\def\picfilename{so8b.eps}
\ifx\nopictures Y\else{\ifx\epsfloaded Y\else\fi
\global\let\epsfloaded=Y
\centerline{\ifx\picnaturalsize N\epsfxsize \picsize\fi
\epsfbox{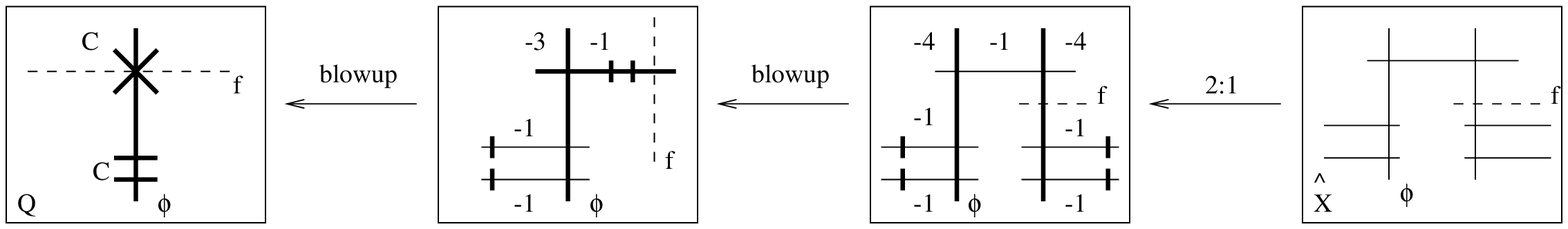}}}\fi
\vskip -.0in

\centerline{Fig.12b: Collision leading to $SO(8)$ ($\Gamma_{0}(2)$ 
fibration)}

\bigskip

As in the $SU(n)$ case,  here $f$ and $\phi$ denote the fibers of the
inherited and the $\Gamma_{0}(2)$ fibrations respectively (or their
projections to $Q$). From the blowup picture we see that the Dynkin
diagram of the combined Kodaira fibers is:

\vskip .2in
\let\picnaturalsize=N
\def\picsize{2in}
\def\picfilename{so8dyn.eps}
\ifx\nopictures Y\else{\ifx\epsfloaded Y\else\fi
\global\let\epsfloaded=Y
\centerline{\ifx\picnaturalsize N\epsfxsize \picsize\fi
\epsfbox{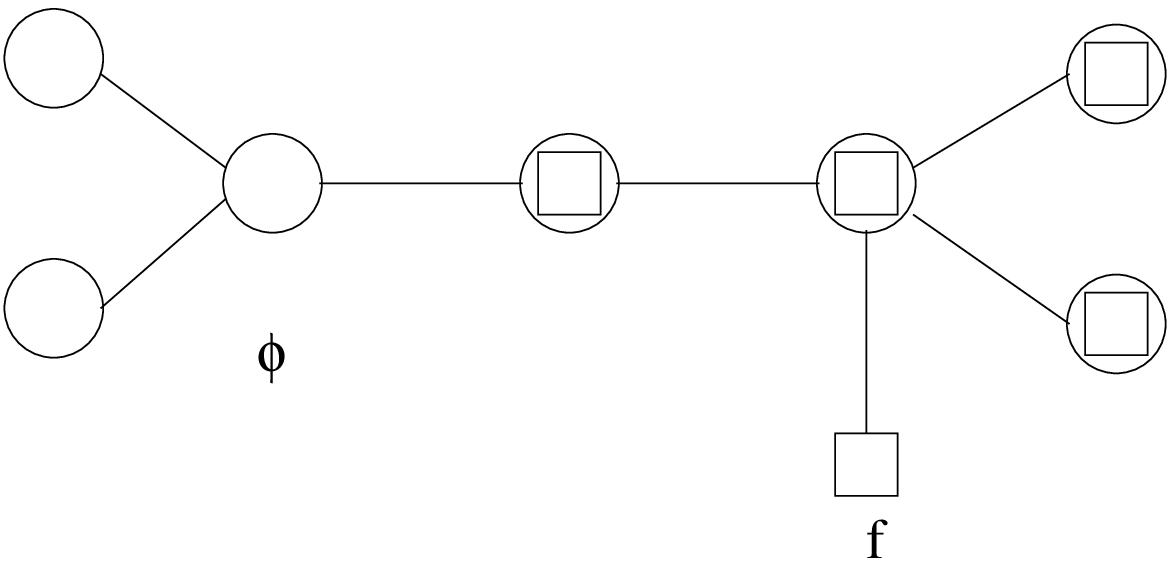}}}\fi
\vskip -.0in

\centerline{Fig.13: Combined Kodaira fibers producing $SO(8)$}

To illustrate the pattern we will work out the next case of a
$2I_{2} + 2I_{1}$ collision in the $\Gamma_{0}(2)$-fibration with the 
two $I_{2}$ coming from different sections. It turns out that this
produces a gauge group $SO(10)$. The blow-up pictures are

\vskip .2in
\let\picnaturalsize=N
\def\picsize{5in}
\def\picfilename{so10a.eps}
\ifx\nopictures Y\else{\ifx\epsfloaded Y\else\fi
\global\let\epsfloaded=Y
\centerline{\ifx\picnaturalsize N\epsfxsize \picsize\fi
\epsfbox{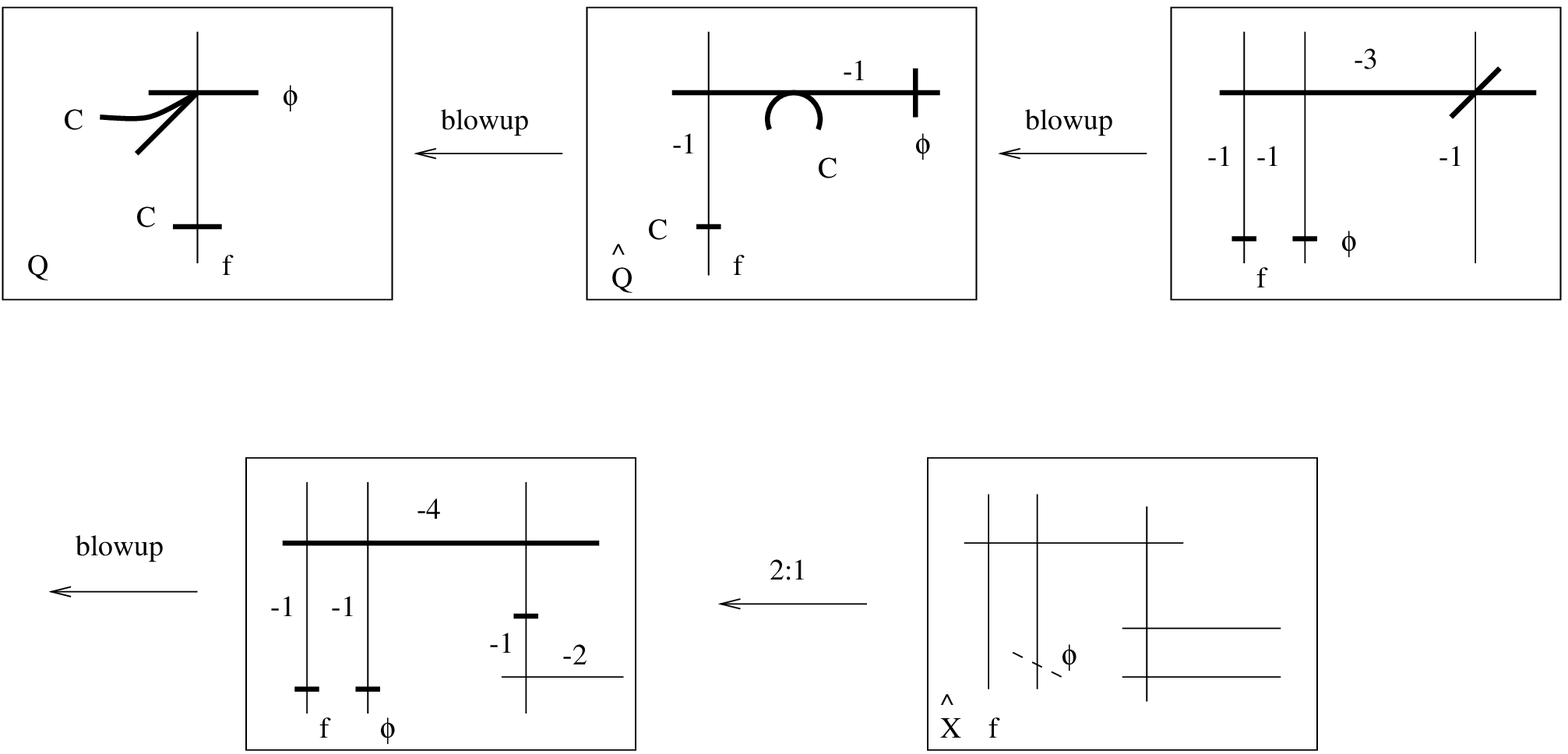}}}\fi
\vskip -.0in

\centerline{Fig.14a: Collision leading to $SO(10)$ 
(inherited fibration)}

\vskip .2in
\let\picnaturalsize=N
\def\picsize{5in}
\def\picfilename{so10b.eps}
\ifx\nopictures Y\else{\ifx\epsfloaded Y\else\fi
\global\let\epsfloaded=Y
\centerline{\ifx\picnaturalsize N\epsfxsize \picsize\fi
\epsfbox{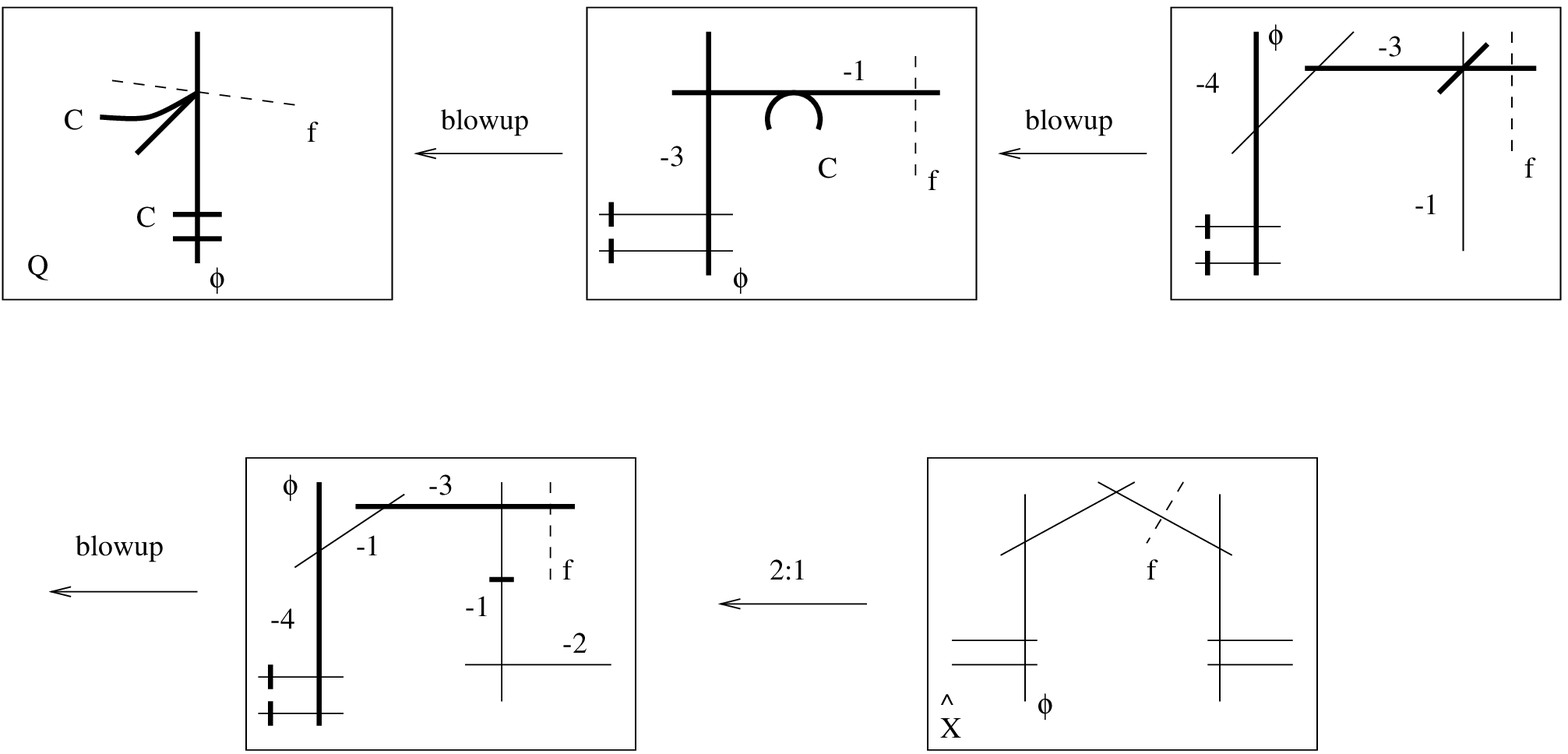}}}\fi
\vskip -.0in

\centerline{Fig.14b: Collision leading to $SO(10)$ ($\Gamma_{0}(2)$ 
fibration)}

\bigskip

and so the combined Dynkin diagram is 

\vskip .2in
\let\picnaturalsize=N
\def\picsize{2.5in}
\def\picfilename{so10dyn.eps}
\ifx\nopictures Y\else{\ifx\epsfloaded Y\else\fi
\global\let\epsfloaded=Y
\centerline{\ifx\picnaturalsize N\epsfxsize \picsize\fi
\epsfbox{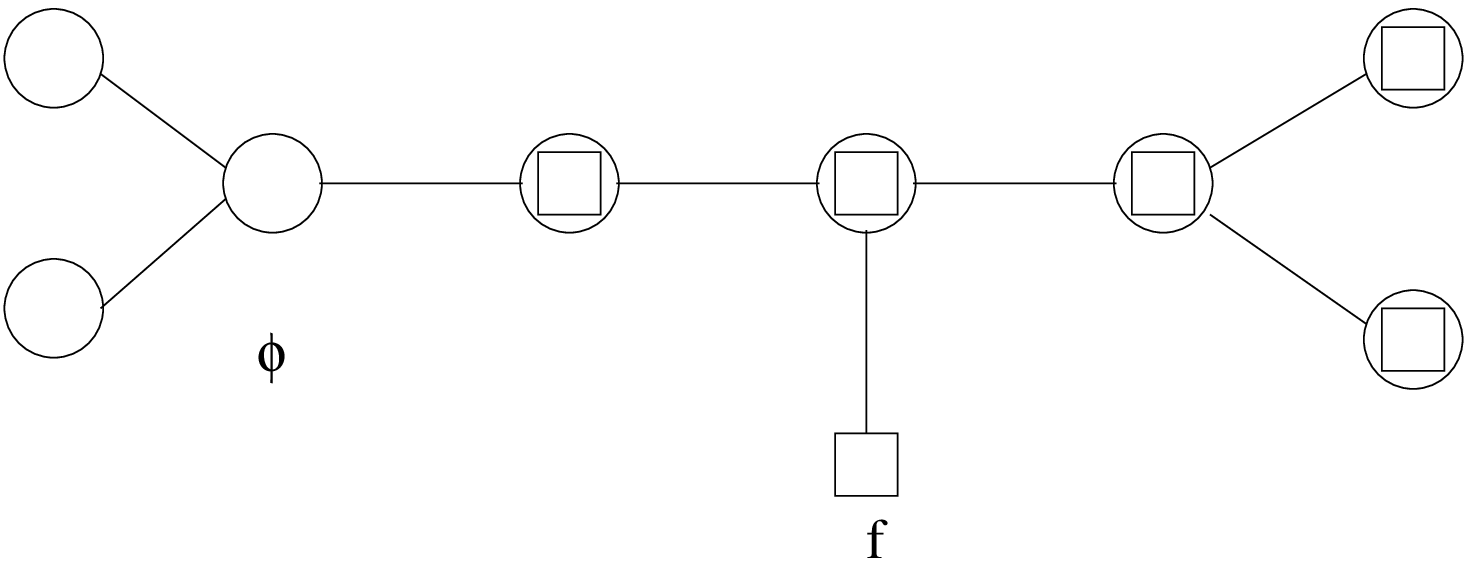}}}\fi
\vskip -.0in

\centerline{Fig.15: Combined Kodaira fibers producing $SO(10)$}

\medskip

\noindent
In general the $\Gamma_{0}(2)$ collision $2I_{2} + nI_{1}$ 
with the two $I_{2}$ coming from different sections leads to a
combined Dynkin diagram

\vskip .2in
\let\picnaturalsize=N
\def\picsize{4in}
\def\picfilename{so2ndyn.eps}
\ifx\nopictures Y\else{\ifx\epsfloaded Y\else\fi
\global\let\epsfloaded=Y
\centerline{\ifx\picnaturalsize N\epsfxsize \picsize\fi
\epsfbox{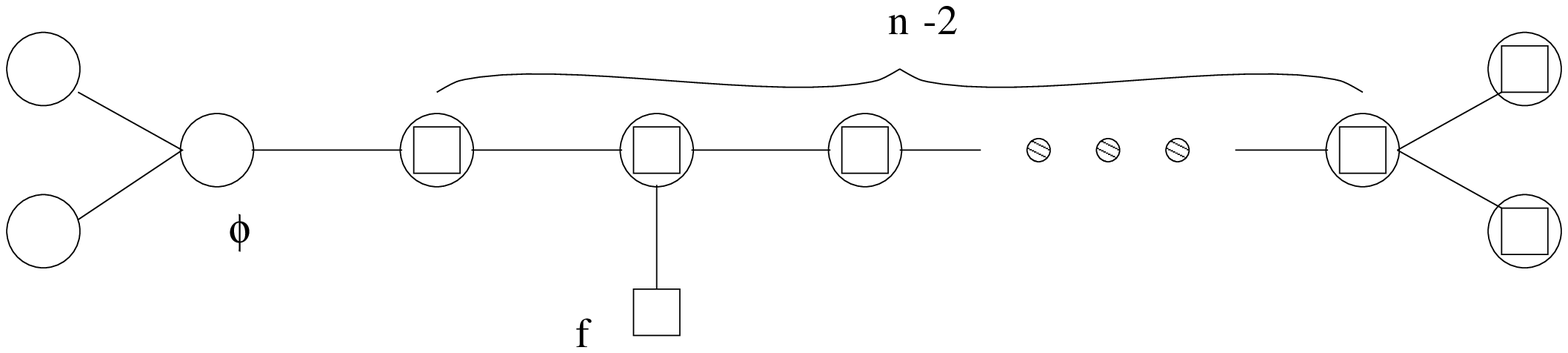}}}\fi
\vskip -.0in

\centerline{Fig.16: Combined Kodaira fibers producing $SO(2n)$}

\medskip

and therefore leads to a gauge group $SO(2n)$.

\bigskip

\bigskip

\noindent
{\it Exceptional gauge groups $E_{6}, E_{7}, E_{8}$:} As in all of the
cases above the appearance of the exceptional groups from the $E_{n}$
series will be governed by the degeneration of the branch curve $C\cup
r_{1}\cup r_{2}$ to a curve with a singularity localized at a single
fiber in either of the fibrations $q_{1}, q_{2} : Q \rightarrow {\bf
P}^{1}$. Similarly to the $SO(2n)$ case the degeneration of $C$ one
needs to get the $E_{n}$ groups is special in the sense that the curve
$C$ becomes {\it globally} reducible. More precisely a component of
bidegree $(1,0)$ has to split of the $(2,4)$ curve $C$. Explicitly  
this means that the branch divisor of $X \rightarrow Q$ is of the form
$r_{1}\cup r_{2}\cup r\cup C'$ where $C'$ is a curve of bidegree
$(1,4)$ and $r$ is a $(1,0)$ curve. Observe that since $C'$ is a 
numerical section of the
fibration $q_{1} : Q \rightarrow {\bf P}^{1}$ we know that $C'$ may be
singular only if it is reducible. 

In terms of the $\Gamma_{0}(2)$ fibration the reducibility of $C$ is
equivalent to having a complete brake-up of the curve of 2-torsion
points, i.e. of geting a monodromy group contained in $\Gamma(2)
\subset \Gamma_{0}(2)$. In other words the $E_{n}$ gauge groups will
come from collisions of fibers in the $\Gamma_{0}(2)$ fibration which
occur in such a way that the monodromy group is kept contained in
$\Gamma(2)$. 

\medskip

\noindent
{\it Gauge group $E_{6}$:} To obtain $E_{6}$ one
needs a $\Gamma(2)$ collision of $6I_{1}$. On the level of the branch
curve this corresponds to having a smooth irreducible $C'$ which is
tangent to third order to $r_{3}$. The relevant blow-up pictures
are shown on Figure~17.

\vskip .2in
\let\picnaturalsize=N
\def\picsize{5in}
\def\picfilename{e6a.eps}
\ifx\nopictures Y\else{\ifx\epsfloaded Y\else\fi
\global\let\epsfloaded=Y
\centerline{\ifx\picnaturalsize N\epsfxsize \picsize\fi
\epsfbox{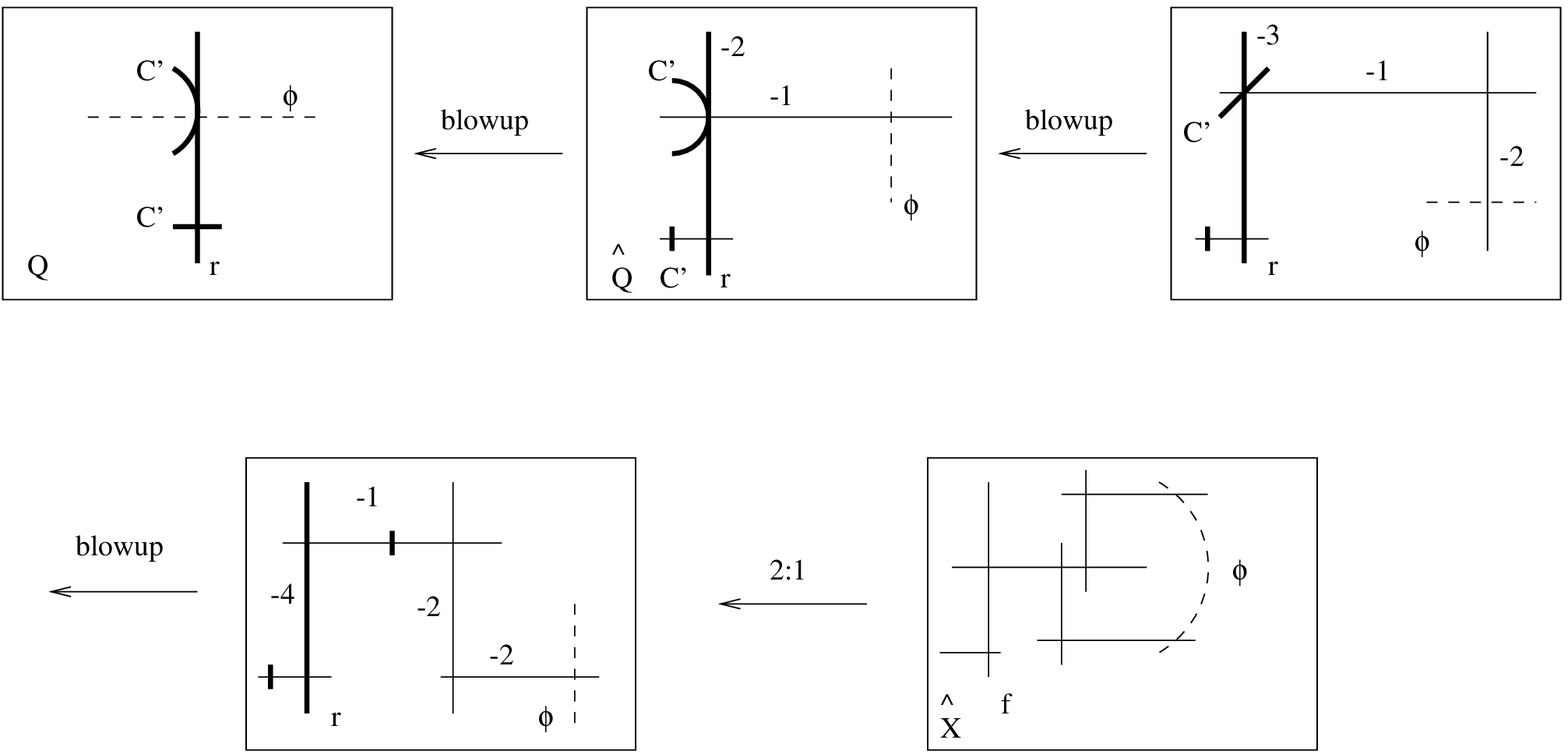}}}\fi
\vskip -.0in

\centerline{Fig.17a: Collision leading to $E_{6}$ 
(inherited fibration)}

\vskip .2in
\let\picnaturalsize=N
\def\picsize{5in}
\def\picfilename{e6b.eps}
\ifx\nopictures Y\else{\ifx\epsfloaded Y\else\fi
\global\let\epsfloaded=Y
\centerline{\ifx\picnaturalsize N\epsfxsize \picsize\fi
\epsfbox{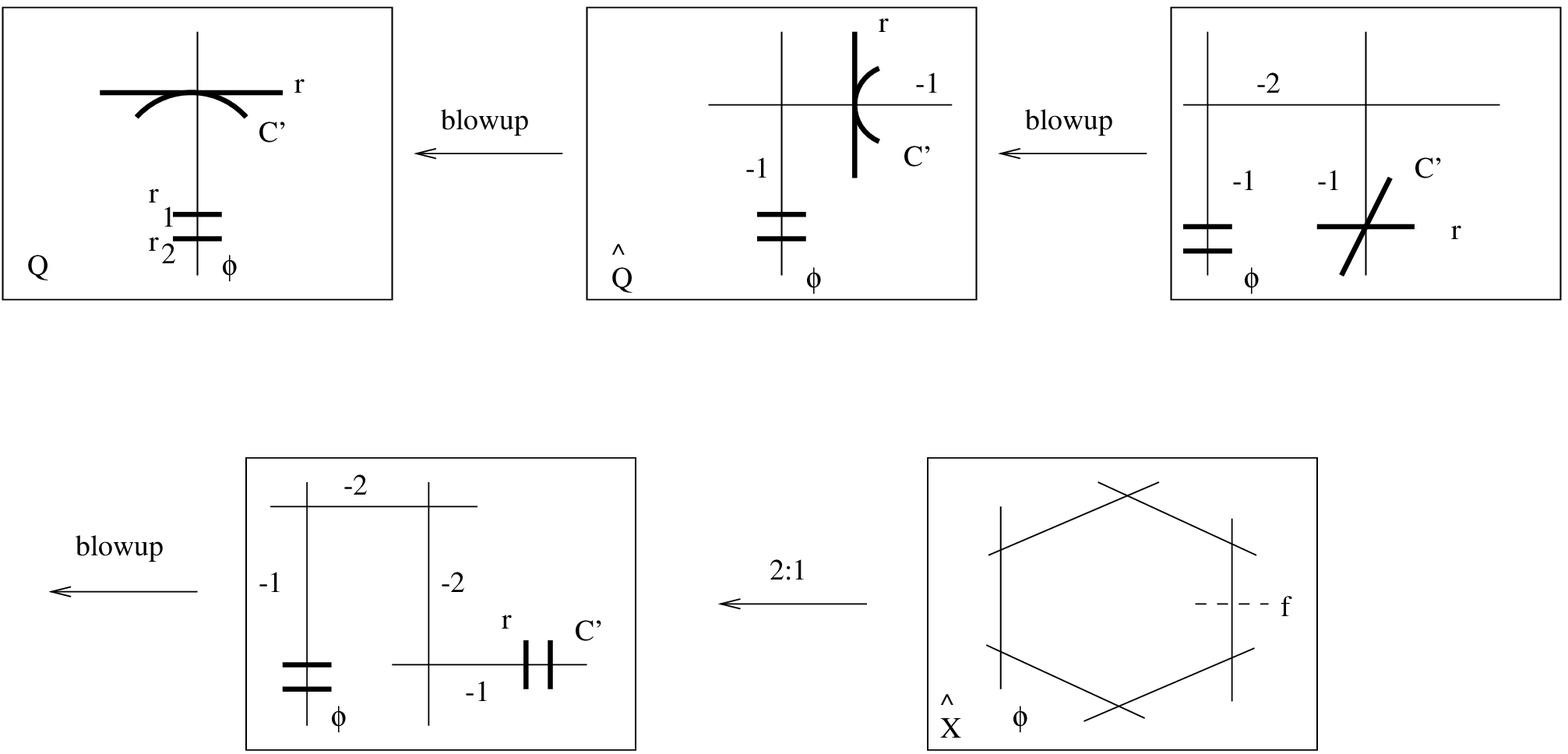}}}\fi
\vskip -.0in

\centerline{Fig.17b: Collision leading to $E_{6}$ ($\Gamma_{0}(2)$ 
fibration)}

\bigskip

\noindent
where as usual $f$ denotes the preimage of the ruling $r$. The
combined Dynkin diagram is:

\vskip .2in
\let\picnaturalsize=N
\def\picsize{2.5in}
\def\picfilename{e6dyn.eps}
\ifx\nopictures Y\else{\ifx\epsfloaded Y\else\fi
\global\let\epsfloaded=Y
\centerline{\ifx\picnaturalsize N\epsfxsize \picsize\fi
\epsfbox{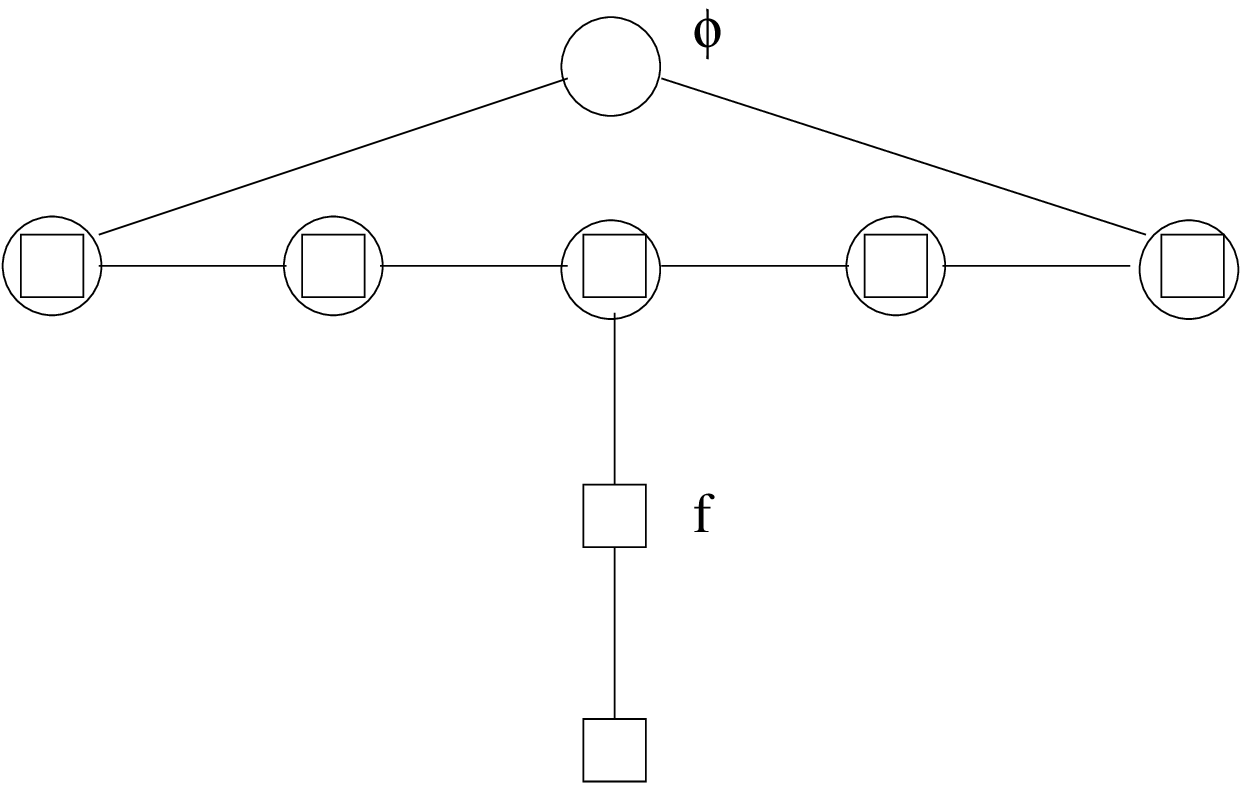}}}\fi
\vskip -.0in

\centerline{Fig.18: Combined Kodaira fibers producing $E_{6}$}

\bigskip

\noindent
Note that in contrast with the cases with $A-C-D$ gauge symmetry
the common part of the two Kodaira fibers does not equal the actual
gauge group (which as before is just the complement of the affine root
of the inherited fiber) but is only a part of it.

\bigskip

\bigskip

\noindent
{\it Gauge group $E_{7}$:} $E_{7}$ can be obtained in two different
ways trough $\Gamma(2)$ collisions. The first one which specializes to
$E_{6}$ occurs when $C'$ becomes tangent to fourth order to $r$,
i.e. in the $\Gamma_{0}(2)$ fibration we have a $\Gamma(2)$ collision
of $8I_{1}$. The necessary blow-ups are shown on Figure~19.

\vskip .2in
\let\picnaturalsize=N
\def\picsize{4in}
\def\picfilename{e76a.eps}
\ifx\nopictures Y\else{\ifx\epsfloaded Y\else\fi
\global\let\epsfloaded=Y
\centerline{\ifx\picnaturalsize N\epsfxsize \picsize\fi
\epsfbox{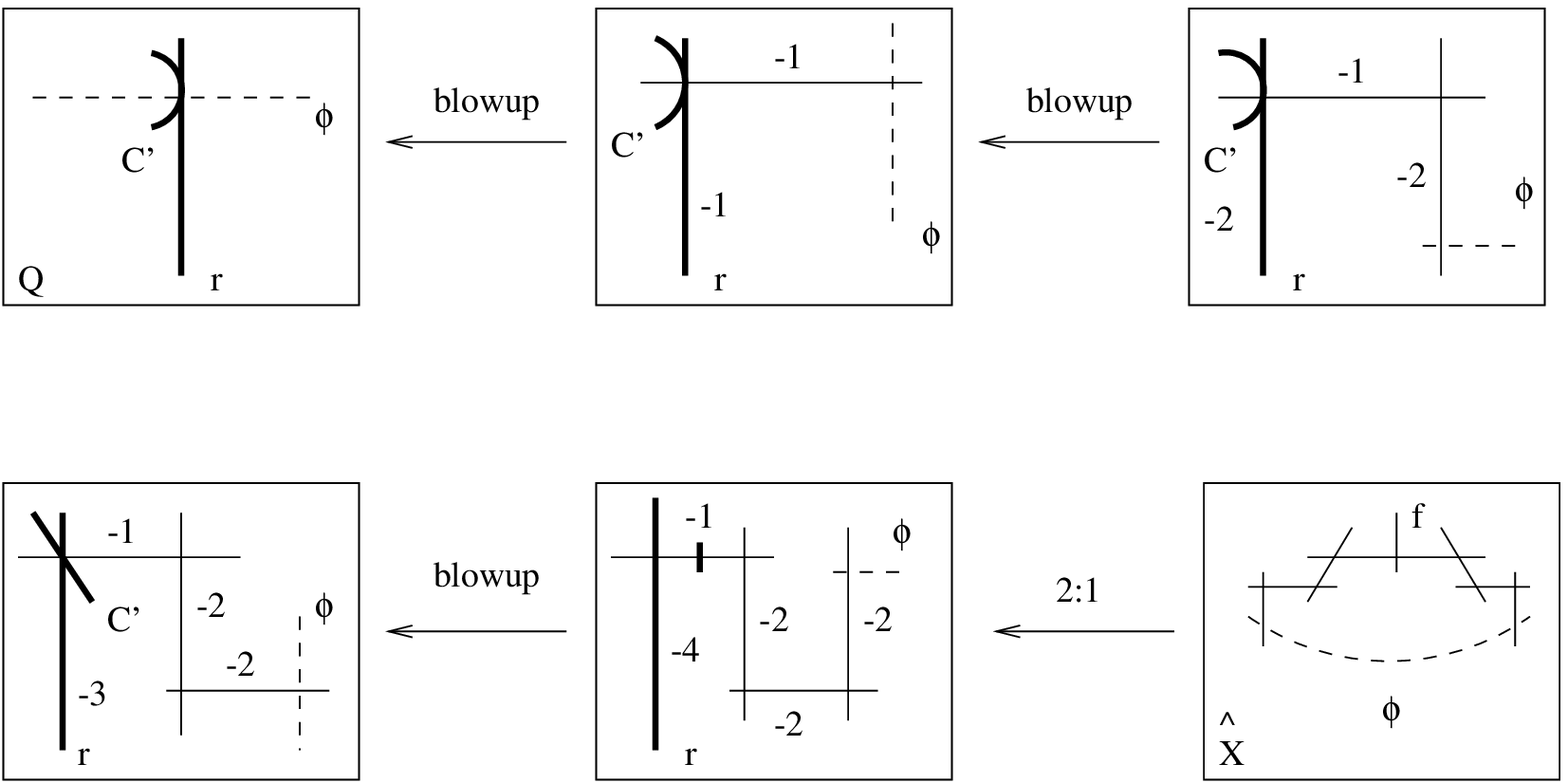}}}\fi
\vskip -.0in

\centerline{Fig.19a: $\Gamma(2)$ collision of $8I_{1}$  leading to $E_{7}$ 
(inherited fibration)}

\vskip .2in
\let\picnaturalsize=N
\def\picsize{4in}
\def\picfilename{e76b.eps}
\ifx\nopictures Y\else{\ifx\epsfloaded Y\else\fi
\global\let\epsfloaded=Y
\centerline{\ifx\picnaturalsize N\epsfxsize \picsize\fi
\epsfbox{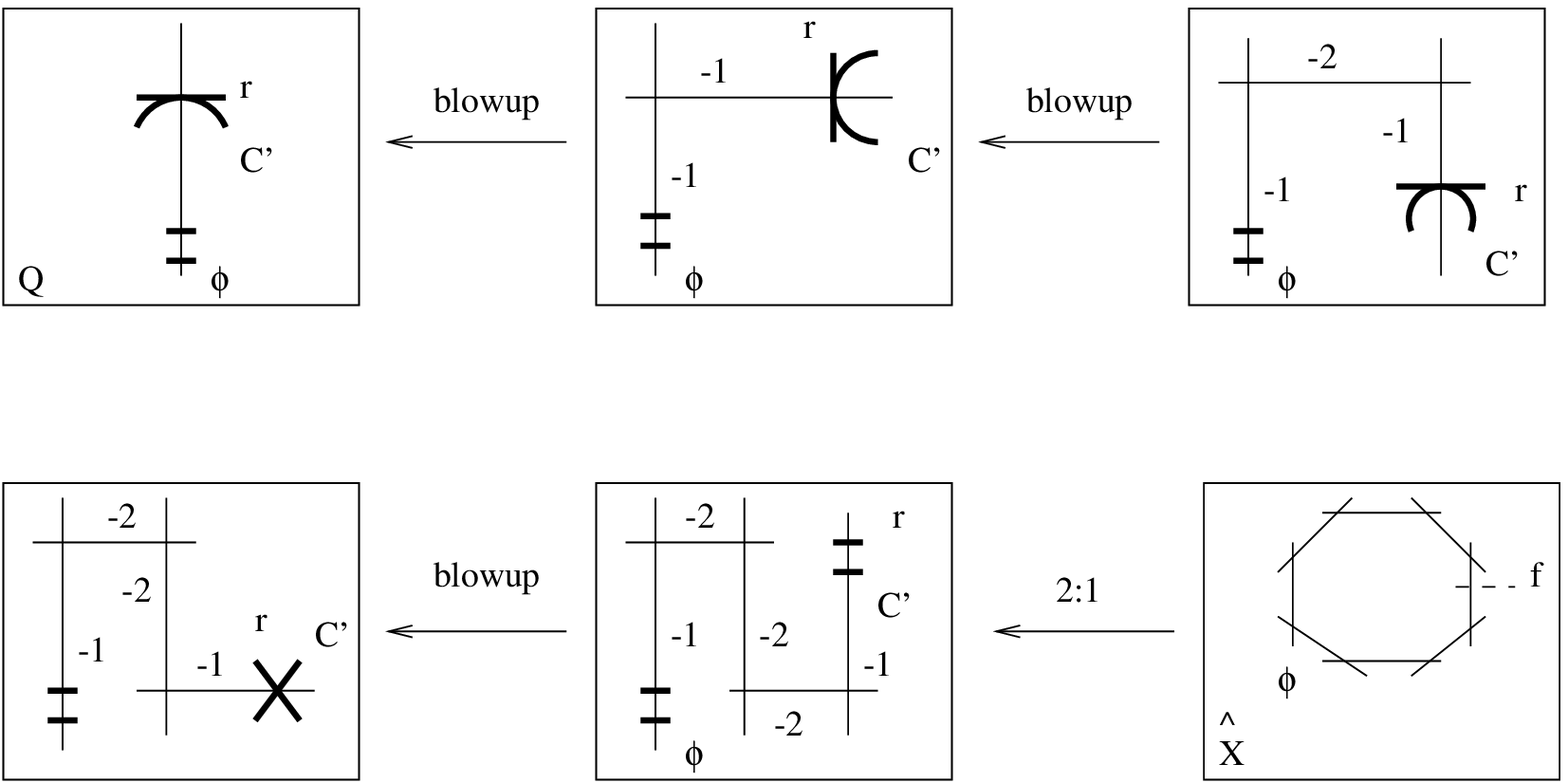}}}\fi
\vskip -.0in

\centerline{Fig.19b: $\Gamma(2)$ collision of $8I_{1}$ leading to
$E_{7}$ ($\Gamma_{0}(2)$ 
fibration)}

\bigskip

\noindent
where $f$ is again  the preimage of the ruling $r$. The
combined Dynkin diagram is:

\vskip .2in
\let\picnaturalsize=N
\def\picsize{3.5in}
\def\picfilename{e76dyn.eps}
\ifx\nopictures Y\else{\ifx\epsfloaded Y\else\fi
\global\let\epsfloaded=Y
\centerline{\ifx\picnaturalsize N\epsfxsize \picsize\fi
\epsfbox{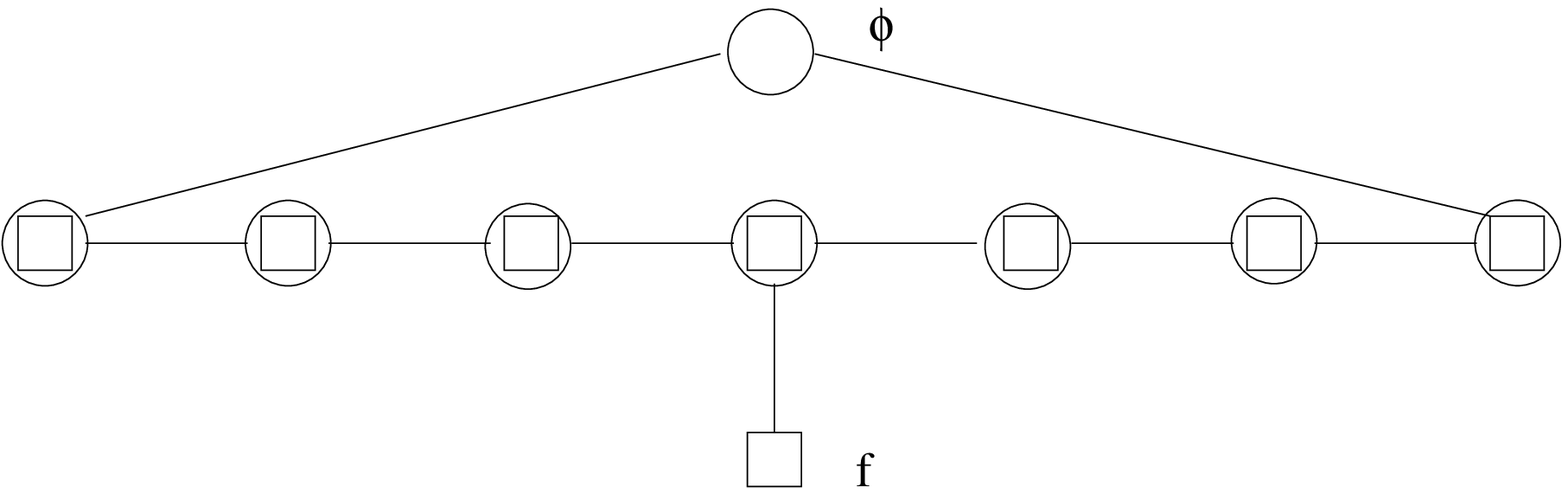}}}\fi
\vskip -.0in

\centerline{Fig.20: Combined Kodaira fibers producing $E_{7}$ in the
$8I_{1}$ collision}

\bigskip

\noindent
The second $\Gamma(2)$ collision giving $E_{7}$ involves another
global break-up. To obtain this collision one needs to find a
degeneration of the configuration $r_{1} \cup r_{2} \cup r \cup C'$ in
which the curve $C'$ becomes singular at a point of $r$. As observed
above this can happen only if a component of bidegree $(0,1)$ splits
of $C'$. This component will necessarily hae to be a fiber of the
projection $q_{2}$ and so we actually have a degeneration of the
branch curve to a curve of the form $r_{1} \cup r_{2} \cup r \cup \phi
\cup C''$ where as before $r_{1}, r_{2}, r$ are fibers for $q_{1}$,
$\phi$ is a fiber of $q_{2}$ and $C''$ is a smooth irreducible curve
of bidegree $(1,3)$. To get $E_{7}$ in this way the curve $C''$ has to
be tangent to $r$ at the point of intersection of $r$ and $\phi$. In
terms of the $\Gamma_{0}(2)$ fibration this corresponds to a
$\Gamma(2)$ collision $2I_{2} + 6I_{1}$ where the two $I_{2}$'s come
from different sections (or equivalently different groups of four in
the Weierstrass model $Y$).

The blow-up sequence producing this $E_{7}$ fiber is shown on
Figure~21.

\vskip .2in
\let\picnaturalsize=N
\def\picsize{4in}
\def\picfilename{e78a.eps}
\ifx\nopictures Y\else{\ifx\epsfloaded Y\else\fi
\global\let\epsfloaded=Y
\centerline{\ifx\picnaturalsize N\epsfxsize \picsize\fi
\epsfbox{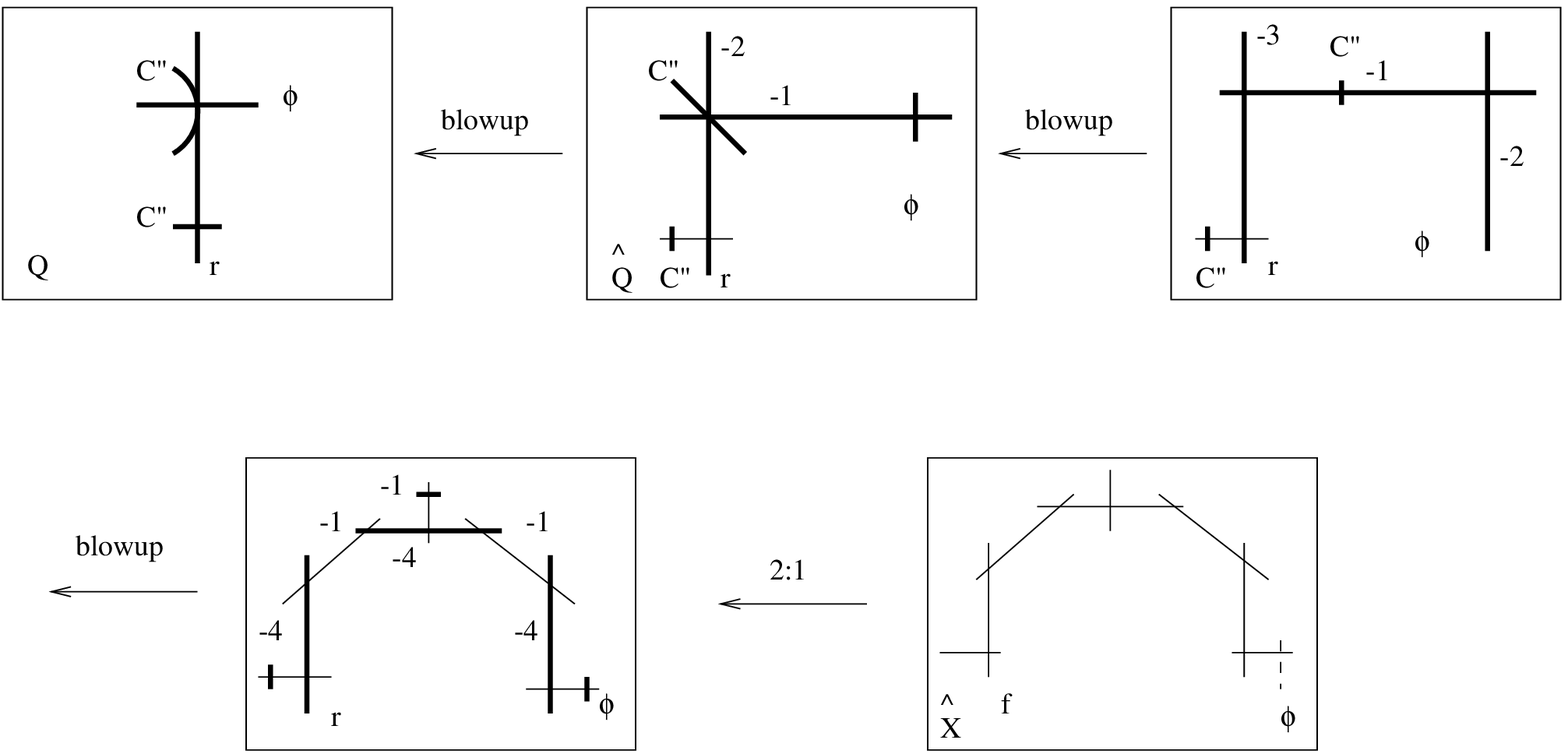}}}\fi
\vskip -.0in

\centerline{Fig.21a: $\Gamma(2)$ collision of $2I_{2} + 6I_{1}$  
leading to $E_{7}$ 
(inherited fibration)}

\vskip .2in
\let\picnaturalsize=N
\def\picsize{4in}
\def\picfilename{e78b.eps}
\ifx\nopictures Y\else{\ifx\epsfloaded Y\else\fi
\global\let\epsfloaded=Y
\centerline{\ifx\picnaturalsize N\epsfxsize \picsize\fi
\epsfbox{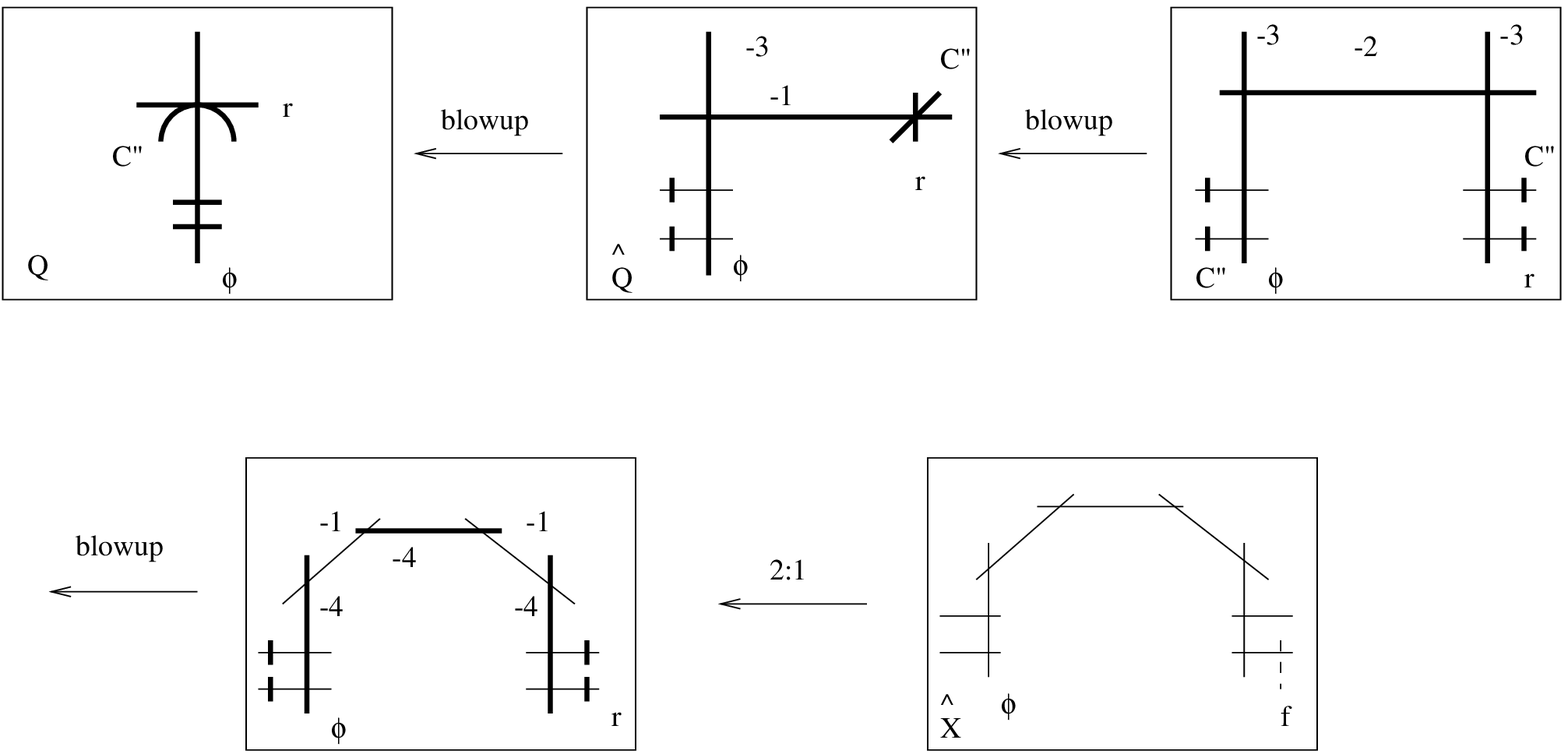}}}\fi
\vskip -.0in

\centerline{Fig.21b: $\Gamma(2)$ collision of $2I_{2} + 6I_{1}$ leading to
$E_{7}$ ($\Gamma_{0}(2)$ 
fibration)}

\bigskip

\noindent
and the combined Dynkin diagram is:

\vskip .2in
\let\picnaturalsize=N
\def\picsize{3.5in}
\def\picfilename{e78dyn.eps}
\ifx\nopictures Y\else{\ifx\epsfloaded Y\else\fi
\global\let\epsfloaded=Y
\centerline{\ifx\picnaturalsize N\epsfxsize \picsize\fi
\epsfbox{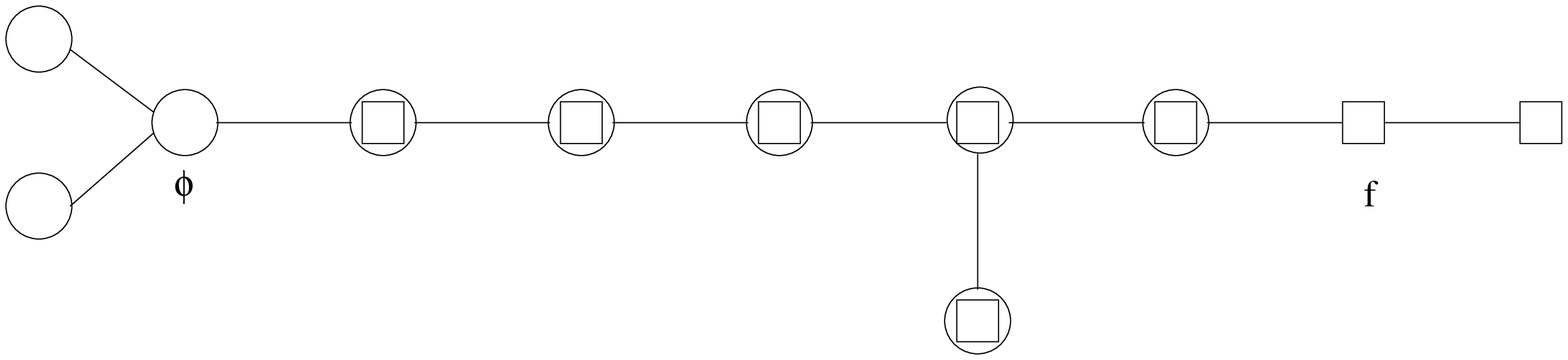}}}\fi
\vskip -.0in

\centerline{Fig.22: Combined Kodaira fibers producing $E_{7}$ in the
$2I_{2} + 6I_{1}$ collision}

\bigskip

\bigskip

\noindent 
{\it Gauge group $E_{8}$:} The gauge group $E_{8}$ appears as a
further degeneration of the $2I_{2} + 6I_{1}$ producing $E_{7}$. More
precisely the appearance of an $E_{8}$ fiber in the inherited
fibration corresponds to a $\Gamma(2)$ collision of $2I_{2} + 8I_{1}$
in the $\Gamma_{0}(2)$ fibration so that again the two $I_{2}$ come
from different sections. The relevant blowup pictures are shown on 
Figure~23.

\vskip .2in
\let\picnaturalsize=N
\def\picsize{4in}
\def\picfilename{e8a.eps}
\ifx\nopictures Y\else{\ifx\epsfloaded Y\else\fi
\global\let\epsfloaded=Y
\centerline{\ifx\picnaturalsize N\epsfxsize \picsize\fi
\epsfbox{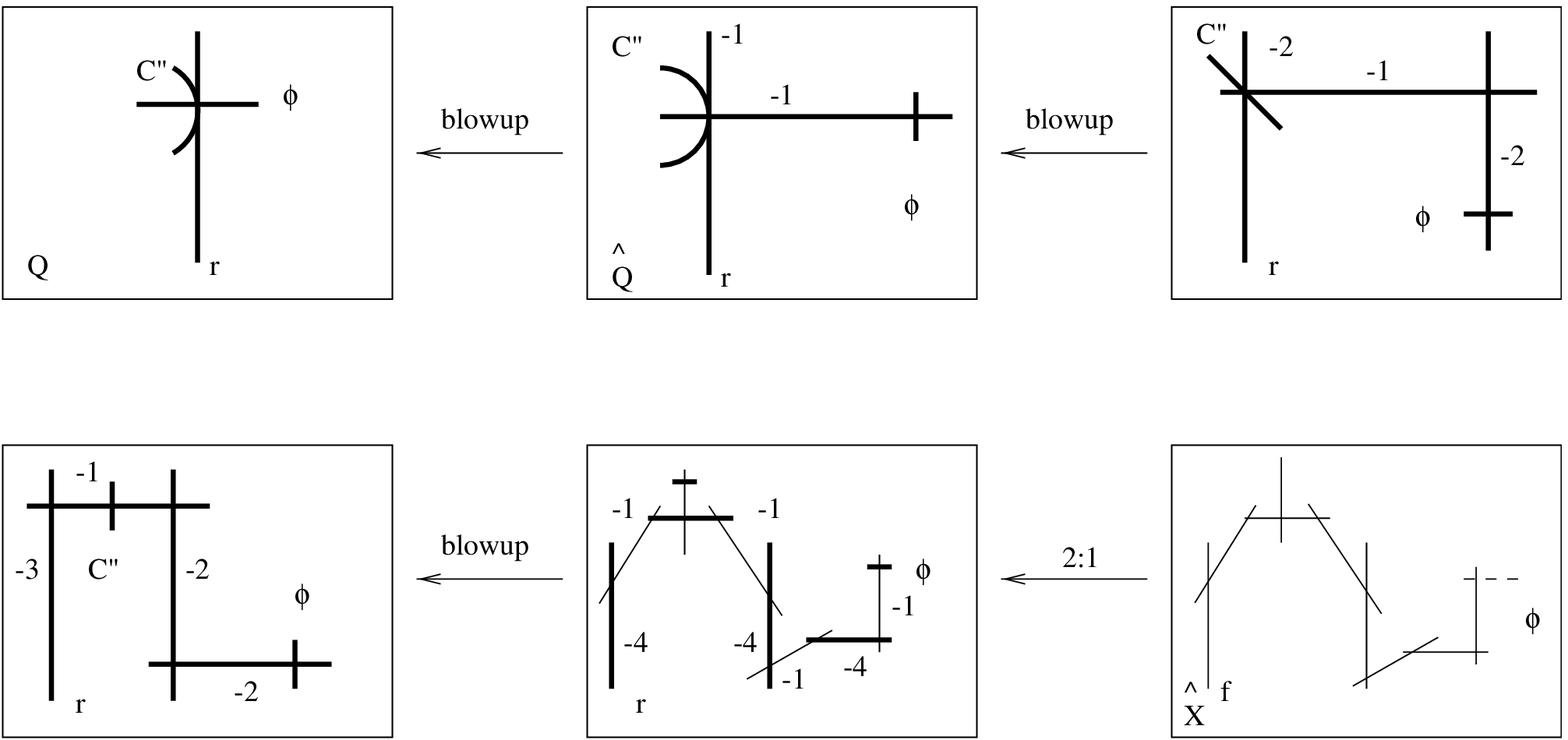}}}\fi
\vskip -.0in

\centerline{Fig.23a: $\Gamma(2)$ collision of $2I_{2} + 8I_{1}$  
leading to $E_{8}$ 
(inherited fibration)}

\vskip .2in
\let\picnaturalsize=N
\def\picsize{4in}
\def\picfilename{e8b.eps}
\ifx\nopictures Y\else{\ifx\epsfloaded Y\else\fi
\global\let\epsfloaded=Y
\centerline{\ifx\picnaturalsize N\epsfxsize \picsize\fi
\epsfbox{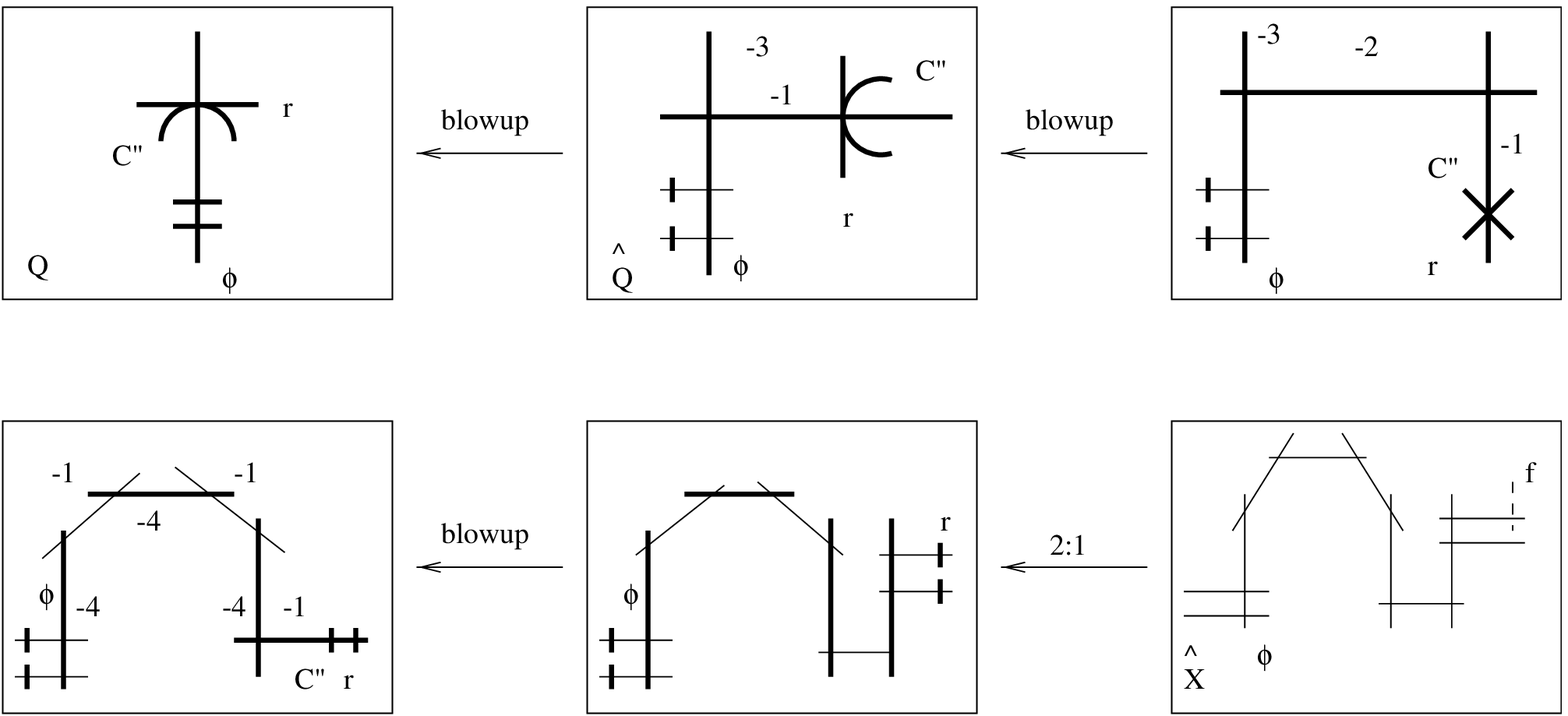}}}\fi
\vskip -.0in

\centerline{Fig.23b: $\Gamma(2)$ collision of $2I_{2} + 8I_{1}$ leading to
$E_{8}$ ($\Gamma_{0}(2)$ 
fibration)}

\bigskip

\noindent
and the combined Dynkin diagram is:

\vskip .2in
\let\picnaturalsize=N
\def\picsize{3.5in}
\def\picfilename{e8dyn.eps}
\ifx\nopictures Y\else{\ifx\epsfloaded Y\else\fi
\global\let\epsfloaded=Y
\centerline{\ifx\picnaturalsize N\epsfxsize \picsize\fi
\epsfbox{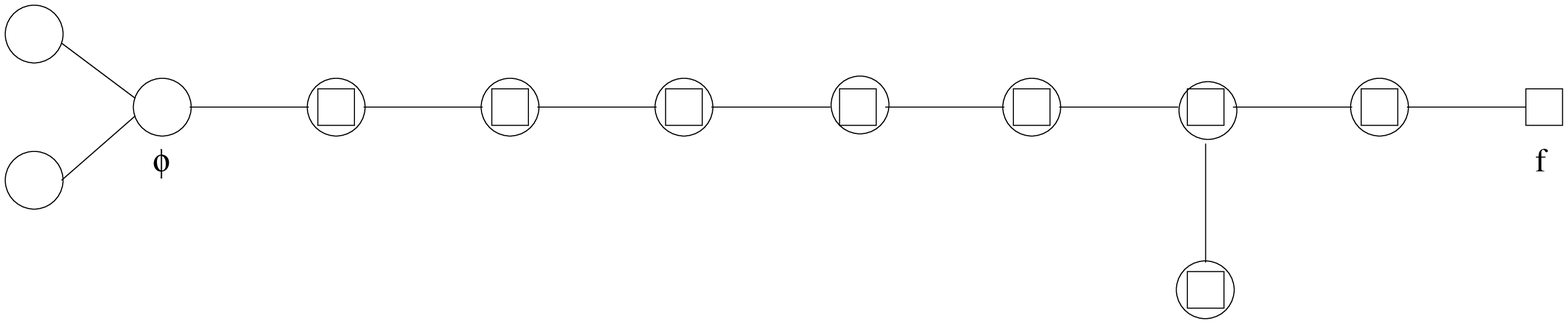}}}\fi
\vskip -.0in

\centerline{Fig.24: Combined Kodaira fibers producing $E_{8}$}

\newsec{Appendix: Review of F-theory/Heterotic duality}

In this appendix we review the features of the F-theory--Heterotic string
correspondence which are instrumental in recognizing the
F-theory dual to a CHL string.

\subsec{The on-the-nose correspondence}

There are several ways in which one can justify this correspondence
The model case is that of eight dimensional compactifications (cf
\vmtwo).  In this case the heterotic string theory
is compactified on a real two
dimensional torus $T^{2}$ and is described by the
following vacua\foot{In this discussion we have suppressed the
heterotic string coupling constant}: a complex
structure $\tau$ on $T^{2}$,  a complexified K\"{a}hler structure
$\rho$ on
$T^{2}$, as well as the Wilson lines of a gauge field on
the torus with structure group $E_{8}\times E_{8}$ (or $Spin(32)/({\bf
Z}/2)$). The F-theory compactification on the other hand takes
place on an
elliptic K3 surface $X$ where the relevant fields\foot{The parameter
corresponding to the heterotic string coupling appears in this context
as the volume of the section of the elliptic fibration.} are the complex
structure on the K3 surface, the elliptic fibration structure $\pi : X
\rightarrow {\bf P}^{1}$ on $X$ and a section
$s : {\bf P}^{1} \rightarrow X$ of the elliptic fibration. The
duality between the two theories is exhibited as an identification of
their moduli spaces of vacua with the locally symmetric space
\eqn\narspace{
O(2,18;{\bf Z})\backslash O(2,18;{\bf R})/(O(2;{\bf R})\times
O(18;{\bf R})).
}
For the heterotic string the identification of the moduli space with
\narspace  \ is given by the Narain construction \ref\narain{K.S. Narain,
{\it New heterotic string theories in uncompactified dimensions
less than 10}, Phys. Let. 169B(1986), 41-46.}.
The identification for the F-theory moduli space is obtained through
the global Torelli theorem for lattice polarized K3
surfaces\foot{Recall \dol that if $M$ is an even
non-degenerate lattice of
signature $(1,\ell)$ a K3 surface $S$ is called $M$-polarized if it is
equipped with an inclusion $M \subset {\rm Pic}(S) \subset
H^{2}(S,{\bf Z})$.}.
Concretely if ${\bf U}$ denotes the two dimensional
hyperbolic lattice and ${\bf E}_{8}$ denotes the root lattice of the
group $E_{8}$ one considers the K3 lattice $\Lambda := {\bf U}^{\oplus
3}\oplus {\bf E}_{8}^{\oplus 2} \cong \Gamma^{3,19}$. It is well
known (see
e.g. \bpv\ )
that this lattice is (non-canonically) isomorphic to the second
integral cohomology of any K3 surface. For any even non-degenerate
lattice  $M \subset \Lambda$ of signature $(1,\ell)$ a version of
the global
Torelli theorem (see the discussion after Proposition 3.3 in \dol) for
K3  surfaces asserts that moduli space
of $M$-polarized K3 surfaces is isomorphic to the space
$\Gamma_{M}\backslash O(M^{\perp}\otimes {\bf
R})/K$. Here $K \subset O(M^{\perp}\otimes {\bf
R})$ is a maximal compact subgroup and  $\Gamma_{M} =
\ker[O(M^{\perp}) \rightarrow Aut((M^{\perp})^{\vee}/M^{\perp})]$.
If $\pi : X \rightarrow {\bf P}^{1}$ is
an elliptic K3 surface with a
section the sublattice in $H^{2}(X,{\bf Z})$ generated by the class
$f$ of the fiber and the class $s$ of the section is isomorphic
to ${\bf U}$ (e.g. via the choice of basis $\{ f , f +
s\}$). By degenerating such a surface to an elliptic K3
surface with a section having two fibers of type $II^{*}$ and four
fibers of type $I_{1}$ one can easily see that the orthogonal
complement
of ${\bf Z}f\oplus {\bf Z}s$ is isomorphic to the lattice
$\Gamma^{2,18}$
and hence if ${\bf U}$ is embedded in $\Lambda = {\bf U}
\oplus {\bf U} \oplus  {\bf U}\oplus {\bf E}_{8} \oplus {\bf E}_{8}$
as the  first summand the moduli space of elliptic K3 surfaces with
a section is naturally identified with the moduli space of
$U$-polarized K3 surfaces. On the other hand $U$ is self-dual and
therefore $\Gamma_{U} = O(U^{\perp}) = O(\Gamma^{2,18})$ which leads
exactly to the locally symmetric space \narspace.

\subsec{The correspondence in the stable limit}

The locally symmetric space \narspace \ is in a natural way a
quasi-projective algebraic variety due to the Bailey-Borel
theorem. However the identifications with the F-theory and the
heterotic string moduli just described are apriori only
complex-analytic since they involve the period map for K3 surfaces and
the complexified K\"{a}hler class $\rho$ respectively. However part of
the identification is algebraic in nature which turns out to be
crucial for the understanding of the six and four dimensional
compactifications of the theories \us, \fmw, \ronictp. The algebraic 
geometry of the
identifications above is best understood (see \vmtwo and \fmw,
section 4.4)
in the limiting regime when the heterotic theory is decompactified by
letting the complexified K\"{a}hler class on the two torus $T^{2}$
become infinitely large. This results in degeneration of
the F-theory K3 to a K3 surface having two elliptic singularities of
type $\tilde{E}_{8}$. The points obtained in this way on the boundary
of the heterotic and F-theory moduli are at infinite distance and so
do not correspond to vacua. On the other hand by understanding the
local geometry of the moduli space near such points gives
non-trivial information about the points in the interior.

Concretely, the surviving part of the
heterotic vacua in the limit  $|\rho| \rightarrow \infty$ is encoded in
the algebraic-geometric data of an elliptic curve $Z$, which is just
the torus $T^{2}$ considered as a complex manifold with the complex
structure $\tau$, and an S-equivalence class\foot{It is necessary to
pass to $S$-equivalence since only the Wilson lines of the
$E_{8}\times E_{8}$ instanton are relevant to the heterotic moduli.} of
semi-stable
$E_{8}\times E_{8}$ principal bundle $V'\times V'' \rightarrow
Z$. By letting the area of $Z$ go to infinity one effectively reduces
the size of the instanton $V'\times V''$ to zero which within the
moduli space of $E_{8}\times E_{8}$ bundles on $Z$ can be viewed as a
deformation of $V'\times V''$ to the trivial
bundle. As explained in \ronictp this deformation
cannot be algebraic in general. On the other hand there is a natural
{\it algebraic} deformation of the S-equivalence class of $V'\times
V''$ to the S-equivalence class of the trivial bundle. For this one
just has to observe that the Looijenga moduli space of $E_{8}$ bundles
on $Z$ is a weighted projective space and so there is a natural line
passing trough every two points. In the following we will be working
with this algebraic family of bundles on $Z$ rather than the family
obtained from letting $|\rho| \to \infty$. This is justified by the
expectation that there should be a global
analytic automorphism of a tubular neighborhood of \narain \ that
sends the two degenerations into each other.  It will be very
interesting to construct such an automorphism explicitly.
 From the F-theory point
of view the deformation of an elliptic K3  $X$ with a section  to
a K3 surface with two elliptic singularities of type $\tilde{E}_{8}$
corresponding to a coordinate of the period going to infinity can
again be replaced by an algebraic degeneration which just uses the
Weierstrass equation of $X$ (see \vmtwo and \fmw, Section 4.4). In other
words by varying the coefficients of the Weierstrass equation we can
find a complex threefold ${\cal X}$ which fibers ${\cal X} \rightarrow
\Delta$ over the unit disk $\Delta = \{ t \in {\bf C} | |t| < 1 \}$
so that all the fibers ${\cal X}_{t}$, $t \neq 0$ are smooth elliptic
K3 surfaces with a section, the fiber ${\cal X}_{0}$ over the point $0
\in \Delta$ is an elliptic K3 with two $\tilde{E}_{8}$ singularities
and there is a point $\delta \in \Delta$ with ${\cal X}_{\delta} \cong
X$.

The algebraic-geometric
description of the F-theory--Heterotic duality in the limit is now
described in two steps. First one replaces (see \fmw, Section 4.4 and
\ref\am{P.Aspinwall, D.Morrison, {\it Point-like Instantons on K3
Orbifolds,} Nucl.Phys. {\bf B503} (1997) 533-564, hep-th/9705104.}) the
singular K3 with two isolated $\tilde{E}_{8}$ singularities by its
stable model which is a normal crossing variety consisting of two del
Pezzo surfaces of type $E_{8}$ glued transversally along an elliptic
curve. To achieve this one performs semi-stable reduction on the
family ${\cal X} \rightarrow \Delta$, i.e. uses a sequence of
blow-ups and blow-downs on ${\cal X}$ and possibly finite base changes
on $\Delta$ in order to replace ${\cal X}_{0}$ by a normal crossing
variety. In the case under consideration the stable reduction
requirest only two blow-ups followed by a blow-down. First one
blows-up the two singular points of ${\cal X}_{0}$ in the threefold
${\cal X}$. The two exceptional divisors $X'$ and $X''$ are rational
elliptic surfaces with a section\foot{Recall that a rational elliptic 
surface with a section is a
surface obtained as the blow-up of ${\bf P}^{2}$ in
$9$ points which are the intersection points of two cubic curves.} and
the strict transform of ${\cal X}_{0}$ is a ${\bf P}^{1}$-bundle over
a copy of the elliptic curve $Z$. By contracting this ${\bf P}^{1}$
bundle onto the curve $Z$ one obtains \am a family of surfaces over
$\Delta$ which is isomorphic to ${\cal X}$ outside of $0 \in \Delta$
and whose central fiber is isomorphic to the transversal gluing
$X'\coprod_{Z} X''$ of $X'$ and $X''$ along $Z$. The rational elliptic
surfaces $X'$ and $X''$ are naturally elliptically fibered (by their
anticanonical linear system) and $Z$ sits in each of them as a fiber
of the elliptic fibration. Since $X'\coprod_{Z} X''$ has also a
section which is a union of two ${\bf P}^{1}$ (each of them in turn a
section of the elliptic fibrations of $X'$ and $X''$ respectively)
meeting at the origin of $Z$ we can blow down these sections to
obtain\foot{Recall that an $E_{8}$ del Pezzo is just the blow up of
${\bf P}^{2}$ at eight points in general position.}
two del Pezzo surfaces $R'$, $R''$ of type $E_{8}$ glued along the
elliptic curve $Z$ siting inside each of them as an anticanonical
section.

The second step needed for the duality in
the limit is to interpret (\fmw, Section 4.1) any pair $Z \subset R$
consisting of an
$E_{8}$ del Pezzo surface $R$ and an elliptic curve $Z$ sitting
inside $R$ as
an anticanonical section, as an S-equivalence class of semistable
$E_{8}$ bundles on $Z$. In a rough outline this interpretation follows
from viewing an $E_{8}$ bundle $V \rightarrow Z$ as a homomorphism
$i_{V} : {\bf E}_{8} \rightarrow \hat{Z}$, $i_{V}(\lambda) :=
V\times_{E_{8}} \lambda$ from the $E_{8}$ lattice
${\bf E}_{8}$ to the dual elliptic curve $\hat{Z}$. Similarly given
the pair $Z \subset R$ one gets a homomorphism ${\rm Pic}(R)
\rightarrow {\rm Pic}(Z)$, $L \mapsto L_{|Z}$. It is well known that
the primitive part $\{ L \in {\rm Pic}(R) | L\cdot K_{R} = 0 \}$ of
the Picard group of an $E_{8}$ del Pezzo is isomorphic to the lattice
${\bf E}_{8}$. On the other hand since $Z$ is a section in
$K_{R}^{-1}$ it follows that for any $L \in {\bf Pic}(R)$ for which
$L\cdot K_{R} = 0$ we wil have $\deg (L_{|Z}) = 0$. Thus the pair
$Z \subset R$ can be viewed as a homomorphism ${\bf E}_{8} \rightarrow
\hat{Z}$ as well.

\newsec{Acknowledgments}

We are grateful to  Oren Bergman, Ron Donagi, 
Andrey Johansen, Zurab Kakushadze, and Cumrun Vafa
for useful discussions. 
The research of M.~B.~and V.~S.~ was partially supported by the
NSF grant PHY-92-18167, the NSF 1994 NYI award and the  DOE 1994 OJI
award. The research of T.P. was supported in part by NSF grant
DMS-9800790.

\listrefs

\end